# HODOR: Shrinking Attack Surface on Node.js via System Call Limitation

Wenya Wang
Shanghai Jiao Tong University
Shanghai, China
ducky_97@sjtu.edu.cn

Xingwei Lin
Ant Group
Hangzhou, China
xwlin.roy@gmail.com

Jingyi Wang*
Zhejiang University
ZJU-Hangzhou Global Scientific and Technological Innovation Center
Hangzhou, China
wangjyee@zju.edu.cn

Wang Gao
Shanghai Jiao Tong University
Shanghai, China
gaowang.sjtu@gmail.com

Dawu Gu*
Shanghai Jiao Tong University
Shanghai, China
dwgu@sjtu.edu.cn

Wei Lv
Ant Group
Hangzhou, China
huaxing.lw@antgroup.com

Jiashui Wang
Zhejiang University, Ant Group
Hangzhou, China
jiashui.wjs@antgroup.com

## ABSTRACT

Node.js applications are becoming more and more widely adopted in the server side, partly due to the convenience of building these applications on top of the runtime provided by popular Node.js engines and the large number of third-party packages provided by the Node Package Management (npm) registry. Node.js provides Node.js applications with system interaction capabilities using system calls. However, such convenience comes with a price, i.e., the attack surface of JavaScript arbitrary code execution (ACE) vulnerabilities is expanded to the system call level.

There lies a noticeable gap between existing protection techniques in the JavaScript code level (either by code debloating or read-write-execute permission restriction) and a targeted defense for emerging critical system call level exploitation. To fill the gap, we design and implement HODOR[1], a lightweight runtime protection system based on enforcing precise system call restrictions when running a Node.js application. HODOR achieved this by addressing several nontrivial technical challenges. First, HODOR requires to construct high-quality call graphs for both the Node.js application (in JavaScript) and its underlying Node.js framework (in JavaScript and C/C++). Specifically, HODOR incorporates several important optimizations in both the JavaScript and C/C++ level to improve the state-of-the-art tools for building more precise call graphs. Then, HODOR creates the main-thread whitelist and the thread-pool whitelist respectively containing the identified necessary system calls based on the call graphs mappings. Finally, with the whitelists, HODOR implements lightweight system call restriction using the Linux kernel feature Secure Computing Mode (seccomp) to shrink the attack surface. We utilize HODOR to protect 168 real-world Node.js applications compromised by arbitrary code/command execution attacks. HODOR could reduce the attack surface to 19.42% on average with negligible runtime overhead (i.e., <3%).

*Corresponding authors: Jingyi Wang and Dawu Gu

[1]HODOR refers to the role of Hodor in the Game of Thrones, who sacrificed his life to resist the attack of the White Walkers by holding the door. We name our tool HODOR to signify its protection of the Node.js applications against malicious payloads attack by restricting the use of system calls.

## 1 INTRODUCTION

Node.js is an open-source, cross-platform JavaScript runtime environment [1], which allows JavaScript code to be executed on the server-side. Currently, some well-known websites use Node.js in their web product, such as Paypal [2], LinkedIn [3], Microsoft [4], and Netflix [5]. The success of Node.js largely owes to the libraries that Node.js depends on. Moreover, Node.js application developers can easily invoke and manage third-party libraries through Node Package Manager (npm) [6]. For example, the `libuv` [7] library provides Node.js with asynchronous I/O capabilities and `V8` [8] library provides Node.js with a JavaScript engine.

These libraries enable the Node.js applications to interact with the low level system kernel, which however may pose severe security threats besides convenience. Specifically, Node.js applications are likely to depend on obsolete third-party libraries with different kinds of security issues, which are expanded into the Node.js ecosystem [9–12]. As observed in [10] , one-quarter of the versions of packages, which make up 19.63% of the npm ecosystem, have dependencies on vulnerable packages. such as gadget chain attacks (prototype pollution attacks) [13, 14], inject-related attacks [15], and supply chain attack [11]. Most of them may lead to ACE attacks [16] ( i.e., arbitrary code execution attacks ). Take the prototype pollution attack as an example. When the attacker pollutes `Object.prototype.someattr` and the application runs the code snippet `eval(someobject.someattr)`, s/he can perform the ACE attacks [13, 17]. When armed with the ability of ACE attacks, attackers can further perform critical operations such as reading private files, creating scheduled tasks, and reversing shells.

Existing protection techniques against the above security implication mostly focus on the JavaScript code, either by code debloating [18, 19] or read-write-execute permission restriction [20]. For instance, `Mininode` [18] and `stubbifier` [19] aim to identify and remove useless JavaScript code to reduce the attack surface so that the attackers cannot easily exploit the compromised third-party modules. `MIR` [20] designs a fine-grained read-write-execute permission model and wraps the privilege reduction over every module based on the function closure mechanism. Although effective in different ways, existing protections targeting the JavaScript code level have several limitations faced with diverse exploitation possibilities. First, they cannot limit the attack surface of arbitrary command execution. For example, when the attacker can pass any malicious data to `exec` method, s/he can execute any command without utilizing any third-party modules or even the JavaScript language. Moreover, existing protections [18–20] require to modify the source code of the Node.js application which may influence the application's normal operation to an uncertain degree.

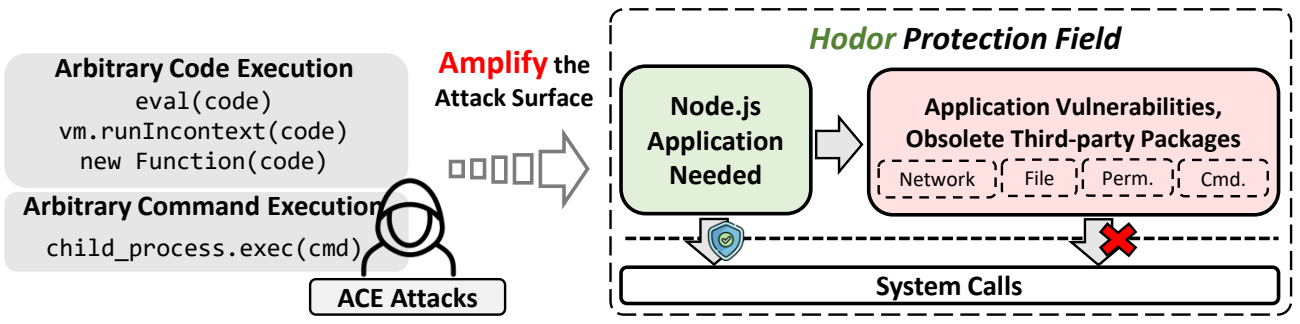

**Figure 1: HODOR protection birdview.**

In this work, we provide a complementary perspective to shrink the attack surface on Node.js applications at the lower system call level to prevent critical operations from attackers as shown in Fig.1. We notice some attack surface reduction works at the system call level for application scenarios other than Node.js [21–29]. However, such low level protection is currently missing for Node.js and several unique challenges are to be properly addressed. ❶ We need to first precisely identify those system calls used by the Node.js application during execution, which require cross-language mapping from the JavaScript source code to the underlying system calls. Previous approaches are not enough as they mostly generate system call lists over a single language [21–28]. Another work `Saphire` [29] scans the AST to identify built-in function calls and builds call graphs over the compiled binary from Debian repositories (to address cross-language challenge), resulting in some loss of accuracy. ❷ Moreover, the highly dynamic nature of JavaScript language makes call graphs construction approaches solely relying on static or dynamic program analysis [18–20] imprecise. ❸ The restriction needs to be transparent to the Node.js framework: the restriction needs to be easily integrated into the Node.js framework and will not influence the normal operation of the application with acceptable overhead. Previous restrictions are mostly performed on the process granularity [21–29], which are inefficient for Node.js which is multi-threaded.

To address the challenges, we develop a lightweight system call level protection system HODOR specifically designed for Node.js applications. HODOR is equipped with three core techniques to tackle each challenge: ❶ and ❷ To generate a precise system call list, HODOR performs cross-language and combined static-dynamic call graphs analysis for the Node.js applications and Node.js framework over JavaScript code and C/C++ code. In particular, HODOR incorporates several significant optimizations to improve state-of-the-art call graphs building methods for both the JavaScript code [30, 31] and the C/C++ code [32]. For JavaScript code, we propose combined static-dynamic analysis to build call graphs for Node.js and we consider the code pattern related to the execution pattern of builtin method features. As for C/C++ code, we propose new partial context-sensitive mechanisms to generate more precise call graphs. Our experimental results show that these optimizations help generate a much more precise set of required system calls for Node.js applications during execution compared to traditional methods. Note that to adapt to the Node.js framework, HODOR distinguishes different system calls executed by different threads. HODOR then create the main-thread whitelist and the thread-pool whitelist respectively containing the identified necessary system calls based on the call graphs mapping. ❸ HODOR finally enforces efficient system call limitation on Node.js with the whitelists using the `seccomp` mechanism in the Linux kernel. Specifically, HODOR implements the restrictions at carefully selected moments to ensure that the protection of different threads does not affect each other.

We implement a prototype of HODOR, and evaluate the effectiveness and efficiency of HODOR in defending 168 real-world Node.js applications compromised by arbitrary code/command execution attacks. The results show that HODOR could reduce the attack surface to 19.42% on average with $< 3\%$ runtime overhead. Overall, we make the following contributions:

- We design and implement a system call level protection system HODOR for Node.js applications. HODOR could accurately infer the system calls used by the applications in the thread granularity at runtime and provide system call restrictions for different threads respectively.
- We propose several significant optimizations in the cross-language analysis to build more precise call graphs for identifying more accurate system call lists. Specifically, we present dynamic/static combined and new **server-side related mechanisms** to improve JavaScript call graphs construction, and new **partial context-sensitive mechanisms** to improve C/C++ call graphs construction.
- We evaluated HODOR by applying it to defend 168 real-world Node.js packages suffering from arbitrary code/command execution attacks. Experimental results confirm the effectiveness and efficiency of HODOR in shrinking the attack surface significantly (**80.57%** reduction on average) with negligible runtime overhead (i.e., <**3%**).
- We released the implementation, evaluation dataset and constructed attack payloads of HODOR[2] to facilitate future research in the Node.js attack and defense area.

## 2 BACKGROUND & MOTIVATION

### 2.1 Node.js Architecture

As shown in Figure 2, Node.js framework consists of builtin module layer, binding module layer, and dependency module layer. Compared with client-side JavaScript runtime, Node.js has two major important features: non-blocking and interaction with OS kernel. We utilize a Node.js application example in Figure 2 to illustrate the working mechanism of Node.js framework. Figure 2 is a code fragment of the version 0.1.0 of `dns-sync` [33] package. By utilizing the method `exec` of the builtin module `child_process`, it could execute the value of `cmd` parameter.

**Builtin module Layer:** Builtin module layer provides Node.js application with builtin modules [34]. Different from JavaScript runtime on browser-side, Node.js runtime provides more functionalities suitable for the server side, such as file system access, asynchronous communication and system command execution. Developers can further build packages on the basis of these builtin modules. Specifically, the application can use the `require` function to invoke builtin modules (Line 1), and these modules use `module.exports` or `exports` to export public API of a module (Line 11).

[2] https://github.com/NodeHodor/Hodor

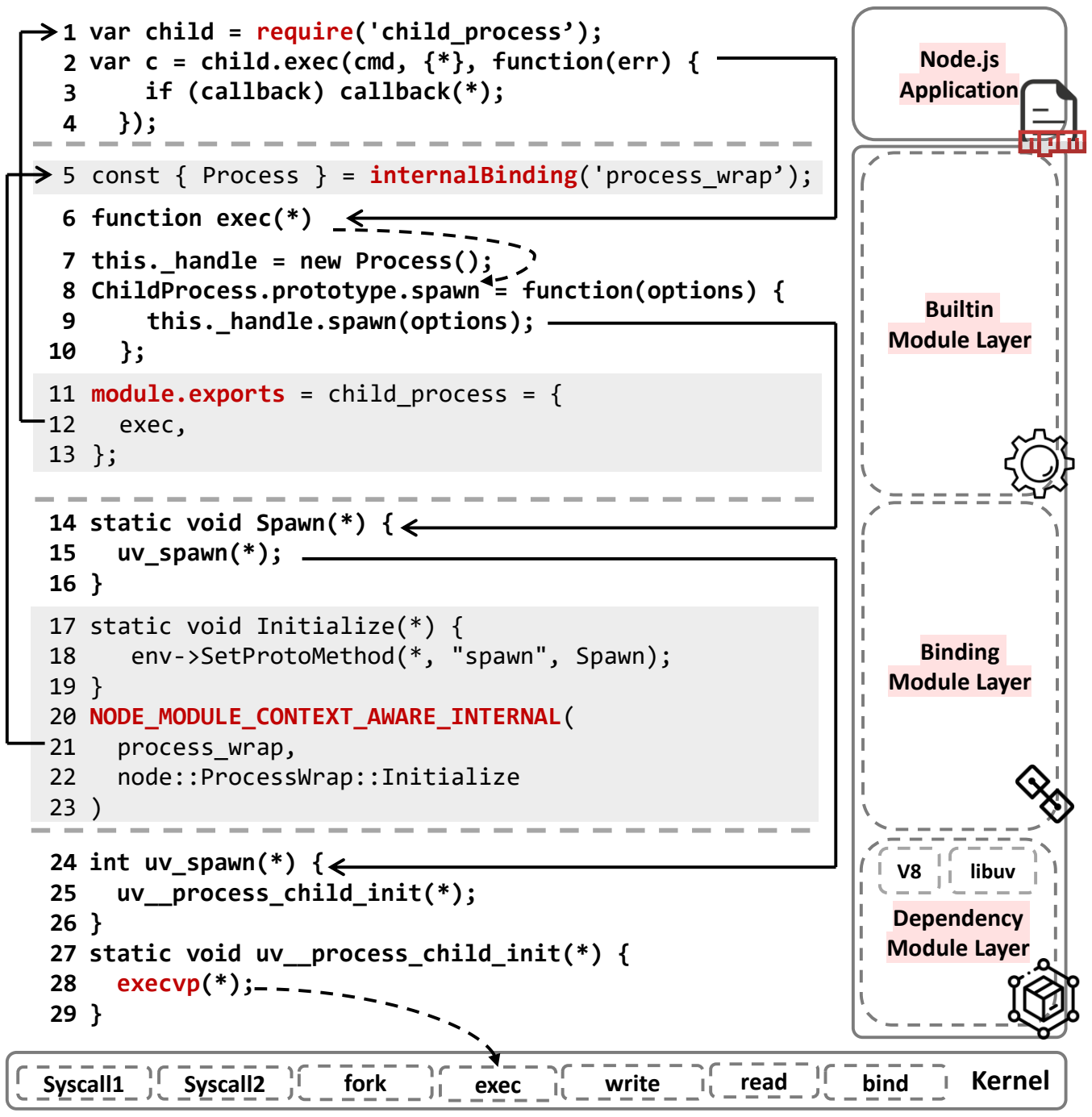

**Figure 2: The overview of Node.js framework.**

**Binding Module Layer:** The binding module layer encapsulates Node.js dependencies as binding modules for builtin modules to invoke. The binding module layer consists of several binding modules. When Node.js engine is running, `RegisterBuiltinModules` will traverse all the `node_module` structures and link them into a global linked list of Node.js, so that the binding modules can be accessed by the builtin module layer through the `internalBinding` method. Each binding module will implement the `Initialize` function (Line 17), and bind the C++ function implemented in this module with the corresponding builtin module layer JavaScript method name. The binding module name and the `Initialize` function will be stored in a `node_module` structure.

**Dependency Module Layer:** Node.js relies on the dependency module layer to achieve this functionality of builtin modules. The dependency module layer consists of several dependencies written in C/C++ code. For example, `V8` provides Node.js with JavaScript bytecode execution environment, `libuv` provides Node.js with non-blocking i/o ability and `OpenSSL` provides Node.js with cryptographic operations. The Node.js application can then use the `internalBinding` function to utilize these dependencies. As shown at Line 5, the builtin modules import builtin module `process_wrap` through `internalBinding` function call.

**Non-Blocking Ability:** As Node.js is non-blocking, it can handle thousands of concurrent connections. Node.js runs in a single process that creates two kinds of threads: *main thread* is used to execute the tasks on the event loop; *threads in the thread pool* are used to handle asynchronous I/O operations. Developers can handle asynchronous I/O operations by calling specific functions. For example, method `readFile` of the module `fs` is executed by the threads in the thread pool and it allows the program to read a file in a non-blocking asynchronous way. And the method `readFileSync` is executed by the main thread and it would block execution until finished. The thread pool would be loaded only when the asynchronous I/O methods are required. Therefore, Node.js applications can be divided into applications that use a thread pool and applications that do not use a thread pool.

## 2.2 Seccomp BPF

`Seccomp BPF` [35] provides a defense mechanism to limit the system call set available to a given application. It used the Berkeley Packet Filter language [36] to allow developers to implement system call filtering with configurable policies. It's worth noting that Seccomp filters could be applied to multi-threaded programs. The threads would inherit the filters of their parents when they are created. In this work, we aim to apply the Seccomp mechanism to limit the system call set of Node.js application and engine at the thread level, in order to implement a thread-aware defense mechanism.

## 2.3 Motivation

We use an example here to illustrate the motivation of our work. Figure 3 shows the `growl` function defined in `growl` library (v1.8.0) with more than three million weekly downloads [37]. In this example, users can use the `growl` function to create a notification message (Line 4). The message is passed to the `exec` method of `child_process` module (Line 15), and is executed as the parameter of `notify-send` command. As the message string is not validated and restricted, command injections can happen. Attackers can inject malicious commands into the message string such as "You have a mail`echo Hacker`", and the *Hacker* string is being printed. **When the attacker is armed with the same permissions as the process that runs the Node.js application, s/he could expand the attack surface and perform critical operations.** For example, she can read the sensitive file (e.g., "mail`cat /etc/passwd`"), create a remote connection (e.g., "mail`nc -l -e /bin/bash 8001`"), and change root permission (e.g., "mail`su root`"). Existing works restrict the attacker's ability at the JavaScript source code level, including application reducer [18] and context wrapper [20]. For attackers who have the same permissions as the process running the application, these protections are not effective.

A recent study [38] evaluatde the packages in the npm repository and observed that 192,585(31.9%) packages in the npm repository do not need access to security relevant resources, such as file systems or network APIs. The inspiration is that **we do not need to grant a running Node.js application access to unnecessarily many system calls**. In this work, HODOR focuses on shrinking the ACE attack surface significantly by system call limitation. For the attack example described above, we identify that the application does not use other system calls except `exec`-related and `notify-send`-related system calls. When running the application, we only allow the application to access `exec`-related and `notify-send`-related system calls while

```
// main.js
var growl = require("growl");
var message = 'You have mail!';
growl(message);

// ./lib/growl.js
exports = module.exports = growl;

var exec = require('child_process').exec
cmd = { pkg: "notify-send" };

function growl(msg, *) {
  args = [cmd.pkg];
  args.push(quote(msg));
  exec(args.join(''),…);
};
```

**Figure 3: An example of ACE attacks.**

disabling its use of file-related, permission-related, and network-related system calls. Therefore, even if an attacker has the ability to execute arbitrary commands, s/he cannot perform many critical operations as s/he wishes.

# 3 THREAT MODEL

We consider the runtime protection of a Node.js application which suffers from two popular types of highly risky attacks: ❶ arbitrary command execution vulnerability where an attacker could execute arbitrary system command, ❷ arbitrary code execution vulnerability where an attacker could execute arbitrary JavaScript code. In practice, the attackers can gain the ability of ACE attacks through injection attack [15], gadget chaining attacks (prototype pollution) [13, 14], and supply chain attack [11]. In general, the functionality of the malicious payloads may include file operation, permission modification, network interaction, and process operation, where a system call is highly likely to be invoked [28].

This work aims to shrink the attack surface of both arbitrary command execution and arbitrary code execution. More precisely, the considered attack surface is the sum of arbitrary code/command execution attack vectors, which can be explicitly quantified as the number of all system calls provided by the system. Our goal is to implement the principle of least privilege for a running Node.js application and minimize the system calls that an attacker can utilize. That is, we only grant a Node.js application access to the minimum set of system calls it requires to run properly. Implementing such system call restrictions does not need additional privileges. It is worth mentioning that our goal is not to prevent ACE attacks but to shrink the attack surface and minimize the critical operations caused by ACE attacks.

This work does not focus on the arbitrary command execution or arbitrary code execution vulnerabilities in the low layer of Node.js (i.e., binding module layer and dependency module layer). Race conditions [39], DOS attacks (such as regular expression DoS) [40, 41], hidden property abuse [42] in Node.js applications, and global variable alterations are also out of the scope of this work, as these attacks are usually triggered without invoking system calls. Install-time attacks [43] is out of the scope of this work since the attacks are triggered during the installation. Moreover, these can be addressed by complementary techniques [20, 39–44].

# 4 SYSTEM DESIGN

The goal of HODOR is to generate system call whitelists for Node.js applications and apply a runtime protection mechanism based on the whitelists to applications to shrink the attack surface of ACE attacks.

## 4.1 Overview

Figure 4 shows the overall framework of HODOR containing four main steps. ❶ We first adopt the **call graph constructor** for both the source code of Node.js application and the Node.js engine. ❷ Next, based on the call graphs, we **build mapping relationships** between the builtin module APIs and their system call list. ❸ Then, by analyzing the call graphs of Node.js application, we identify the builtin modules used by the Node.js application. And based on the mappings between methods of builtin modules and system calls, we **generate whitelists** (of system calls) for the application. ❹ Finally, we apply the **whitelist restriction** of system calls protection mechanism to the threads on which the Node.js application runs.

## 4.2 Call Graph Constructor

The system calls used by a Node.js application are encapsulated layer by layer in Node.js engine. In order to generate a system call lists for a Node.js application, we improve the precision of traditional call graphs constructors (for both JavaScript and C/C++ language) with multiple optimizations, and construct modular call graphs for both the application and the dependent Node.js engine.

**JavaScript Language:** For Node.js applications and builtin modules written in JavaScript, we build their modular call graphs respectively. The state-of-art JavaScript call graphs constructors [30, 31] do not fully consider the execution pattern of the methods in the builtin modules of JavaScript (e.g., `Promise`) and builtin modules of Node.js (e.g., `fs`), as the source code of builtin methods is not included in the applications. However, the execution of these functions triggers the invocation of functions or the creation of objects that appear in their arguments. For example, as shown in Figure 2 on Line 2, the execution of builtin method `child_process.exec` triggers the execution of the callback function in the last argument, and the execution of builtin method `Function.binds` creates a new function. Not considering such cases can result in a significant number of missing edges from the callback function invocation to the callback function definition of the function nodes creations. In this work, we summarize the execution pattern of the builtin methods provided by JavaScript [45] and Node.js [34] and add the function nodes and edges related to the function execution.

The highly dynamic nature of JavaScript code leads to unsound call graph construction [30]. For example, the JavaScript code can utilize dynamic constructors to create new functions represented as a string or execute JavaScript code represented as a string and can perform dynamic addition and deletion of object properties. These cannot be accurately analyzed by static call graph construction and lead to a significant number of missing edges from the function invocation represented as a string to the function definition represented as a string. In this work, we utilize dynamic call graph construction to identify the missing nodes and missing edges and generate combined static-dynamic call graphs for Node.js applications.

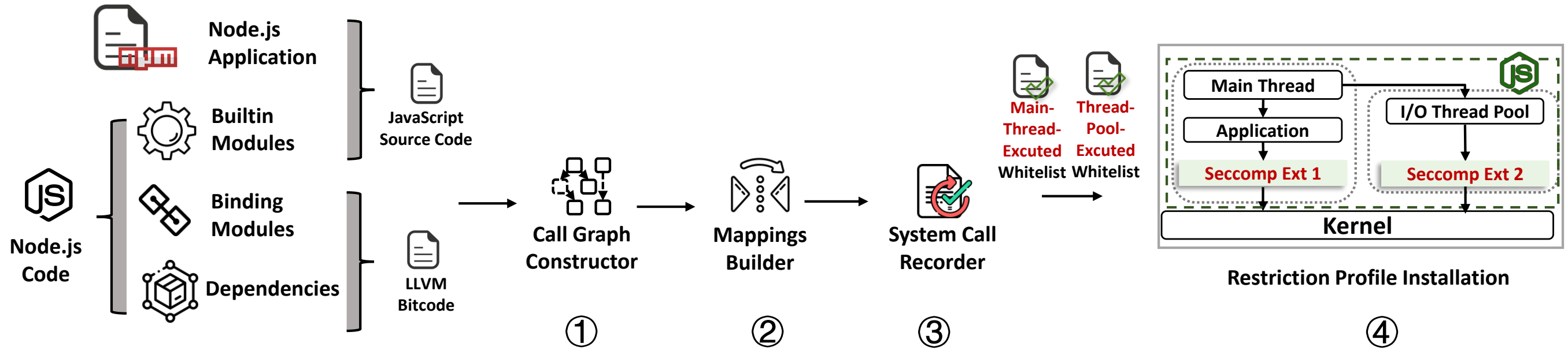

**Figure 4: The HODOR pipeline.**

**C/C++ Language:** When analyzing Node.js binding and dependency modules written in C/C++ language, the traditional context-insensitive call graphs construction method brings considerable false positives. Using this result directly will make the subsequent syscall whitelist overly broad. However, conducting context-sensitive analysis for all the source code will bring extremely large memory and time overhead. Therefore, we propose a **partial context-sensitive** call graphs construction algorithm to handle the above challenges.

We summarize two code patterns leading to false positives and implement context-sensitive analysis for these patterns. The first pattern is **context-insensitive for switch-case statement**. Taking Figure 5 as an example, `uv_fs_access` (Line 3) and `uv_fs_write` (Line 7) call the same `uv__fs_work` function (Line 14). `uv__fs_work` implements diverse logic using switch-case statements and only one case-branch can be reached depending on the function parameters. The context-insensitive algorithm considers all the branches of the switch-case statement to be reached, which causes false positives. We locate functions like `uv__fs_work`, collect the context information on each callsite, and eliminate unreachable case-branches of these functions. The second pattern is **context-insensitive for function-pointer parameter**. Taking Figure 6 as an example, `Read`(Line 1), `Unlink` and `RMDir` functions call `Call` (Line 20) functions through different call chains. The function pointer `fn` will be called finally, and its value changes depending on different call chains. The context-insensitive algorithm considers all the possible functions (like `uv_fs_read`, `uv_fs_unlink` and `uv_fs_rmdir`) to be called through `fn`, which causes false positive. We collect all the functions that call function pointers and perform a backward data-flow analysis to determine the assignment location of function pointers and record the call chains. We traverse the call chain and clone each function in it to make sure that every different value of the function pointer pass to different cloned functions (like `Call`).

## 4.3 Mappings Builder

After obtaining call graphs of each layer, we identify the function call relationships between layers at this stage. We finally generate a mapping set between builtin modules and system calls.

**Builtin Module Layer:** Starting with the builtin methods as entry points, we traverse the call graphs of builtin module layer and collect the binding methods used by builtin methods. As shown in Figure 2, we generate the mapping between builtin method `exec` method on Line 12 and the `spawn` method of binding method `spawn` on Line 9.

**Binding Module Layer:** Referring to the code pattern of registering the C++ modules and methods, we implement an LLVM Pass to collect the mappings between binding methods and C++ functions. As shown in Figure 2, we record the binding module `process_wrap` and its method `spawn` on Line 20. Furthermore, on Line 18, we generate the mapping between the method and C++ function `Spawn` defined on Line 14.

**Dependency Module Layer:** We generate the mapping between C/C++ functions and `libc` functions based on the call graphs of the binding module layer and dependency module layer. For example, we generate the mapping between C++ function `uv_spawn` on Line 24 and the `libc` function `execvp` on Line 28 in Figure 2.

Finally, combining the mappings between `libc` functions and system calls, with the mappings of three layers, we build the mappings between builtin modules and system calls.

## 4.4 Whitelist Generator

At this stage, we generate system call whitelists for applications. We first identify which builtin modules and builtin methods are used by the application at runtime based on the call graphs generated in Section 4.2. Furthermore, based on the mapping relationships generated in Section 4.3, we link the builtin modules and methods with their system call whitelist. As discussed in Section 2.1, the tasks of the main thread and the thread pool are different, and the required system calls are also different. Therefore, to achieve a thread granularity protection mechanism, we divide the system call whitelist into **main-thread-executed** whitelist and **thread-pool-executed** whitelist based on the code pattern of two kinds of threads.

## 4.5 HODOR Builder

We generate the corresponding system call filters based on system call whitelists and use the `seccomp` facility to restrict the actions available within the threads.

A challenge in loading the filters is the timing of loading the filters for the main thread and the child threads in the thread pool separately. Child threads in the thread pool will only be created by the main thread when the tasks arrive. Considering the inheritance feature of `seccomp`, i.e., the child threads in the thread pool inherit filters of the parent (main) thread when they are created by the parent (main) thread. However, the main thread and the thread pool perform different tasks, inherited filters will block the execution of child threads in the tread pool. To overcome this challenge, we design a fine-grained filtering mechanism, where for different types of applications, we adopt different loading mechanisms. Especially, for thread pool required applications, we add the filters at two carefully selected moments. We illustrate the detail implementation in Section 5.5.1.

# 5 IMPLEMENTATION

In this section, we elaborate the core implementation details of HODOR to realize the design goals. To generate a more complete system call list for Node.js applications, we adopt static analysis and dynamic analysis for Node.js applications and Node.js engine.

## 5.1 JavaScript Call Graph Construction

We first perform static analysis for JavaScript code. Referring to `JAM` [30], we build modular call graphs for Node.js applications and Node.js builtin modules. The call graph constructor also considers JavaScript promise-chain [46] and execution pattern of builtin methods [34, 45]. `JAM` is the state-of-art modular call graphs building methods for Node.js applications. However, we could not access the source code of `JAM`. We reproduce the methodology of constructing modular JavaScript call graphs based on the rules of `JAM`.

Moreover, different from the previous work, our work also considers the execution pattern of builtin methods of Node.js engine and JavaScript. `JAM` [30] and `js-callgraph` [31] did not record the functions that act as a parameter of the builtin method. For Node.js builtin methods, functions can be executed as a callback function. For example, as shown in Figure 2, the anonymous function on Line 2 will be invoked after executing the command. In our work, we first manually collect 32 builtin methods whose parameters are callback functions and record the location of the callback function based on the Node.js API document [34]. Then, when building call graphs, we match these builtin methods and add the edge pointing from the domain where such a method is invoked to the point where the callback function is defined. The state-of-art call graphs building tools also omit the execution pattern of builtin methods of JavaScript. For example, when the function wrapped by the `Promise` object is finished, the callback function of `then` method will be invoked. And when a builtin method `map` of `Array` Object is invoked (e.g., `[1,2,3].map(x => x * 2)`), its first argument would be executed as a function. Moreover, different from builtin methods provided by Node.js engine, developers could also invoke builtin methods (e.g., `Object.create`) to create objects, which would also affect the construction of the call graphs. The method properties of the object may be invoked. In our work, we refer to the Standard built-in objects [45]. We collect ten builtin methods with function parameters including the methods of `Promise` and `Array`. When building call graphs, we match these builtin methods and add the edges pointing from the domain where such a method is invoked to the point where the callback function is defined. Moreover, we also record the nine builtin methods including the methods of `Function` and `Objects` and add nodes when these methods are invoked.

*5.1.1 Dynamic Analysis Refiner.* The highly dynamic nature of JavaScript leads to unsound call graph construction [30]. JavaScript code can use `eval`, `new Function`, and the method of `vm` to execute JavaScript code represented as a string and returns its completion value. For example, `eval("console.log('hello')")` prints hello string, and `let sum = new Function('a','b','return a+b');` creates a function that sums two arguments. However, existing work generally does not analyze the string of these functions accurately, thus the definition of a new function or function call represented as a string cannot be included in the call graphs. To solve this deficiency, we introduce dynamic call graphs analysis. We use the dynamic call graphs tool `Nodeprof` [47]. We first collect the test suites from the packages installed by `npm` or `github.com` and build test suites for the application manually. Then we execute test suites and build dynamic call graphs for the application through the dynamic call graphs construction tool. Finally, we combine the nodes and edges of call graphs conducted by the static analysis and call graphs conducted by the dynamic analysis and generate the combined static-dynamic call graphs for the application.

*5.1.2 Dynamic Command Execution.* Node.js applications can utilize `exec` and `spawn` methods of `child_process` module to execute dynamic commands. For example, as shown in Figure3, `notify-send` command also requires system calls. However, existing protection work does not consider the system calls needed by the commands. Disallowing the execution of the system calls required by the commands can lead to program failure. In this work, we first identify the location of command execution methods at the static analysis phase. We then extract the commands executed by the command execution methods at the dynamic analysis phase. Then we use Linux `strace` utility [48] to record system calls utilized by the commands.

## 5.2 C/C++ Call Graph Construction

The C/C++ call graphs construction method of HODOR is based on `SVF` [32], which is a static program analysis tool for LLVM-based languages. We wrapped `clang` [49] with `wllvm`[50] to compile all the Node.js engine C/C++ source code to LLVM bitcode. By monitoring the command outputs of the compilation process, we utilize `llvm-link` [51] to link all the source code of binding and dependency modules into a single LLVM bitcode file as our call graphs analysis target.

As described in Section 4.2, the traditional context-insensitive call graphs construction algorithm leads to considerable false positives for switch-case statements and function-pointer parameters. `SVF` conducts its default call graphs construction algorithm based on context-insensitive analysis. To balance precision and the overhead of context-sensitive analysis, we design a partial context-sensitive algorithm to eliminate false positive cases.

*5.2.1 Unreachable Branch Elimination for Switch-Case Statement.* We summarize the code patterns to locate functions for switch-case statement context-sensitive analysis: ❶ There exists a switch-case statement in the function and different branches of the switch-case statement would call different functions; ❷ The parameters of the function influence the branch selection variable.

For each call-site of functions that match the pattern: ❶ Performing intra-procedural data-flow analysis to record the parameter or the member field offset within the structure parameter that propagates to the branch selection variable of the switch-case statement. ❷ Performing backward data-flow analysis from the caller function to locate the assignment of the above variables. If the assigned constant can be determined, $\langle \mathtt{func}_{\mathtt{caller}}, \mathtt{func}_{\mathtt{switch-case}},$

```
#define INIT(subtype) req->fs_type = UV_FS_ ## subtype;

int uv_fs_access(*) {
  INIT(ACCESS);
  POST;
}
int uv_fs_write(*) {
  INIT(WRITE);
  POST;
}

#define POST uv__fs_work(&req->work_req);

static void uv__fs_work(struct uv__work* w) { uv__fs_work.uv_fs_access
  req = container_of(w, uv_fs_t, work_req);
  switch (req->fs_type) {
    X(ACCESS, access(req->path, req->flags));
    X(WRITE, uv__fs_write_all(req));
    …
   default: abort();
    }
}
```

**Figure 5: Switch-case statement context-insensitive.**

$constant_{switch-case}\rangle$ would be recorded and $func_{switch-case}$ would be cloned using LLVM. ❸ For each cloned $func_{switch-case}$, case-branches except that belonging to $constant_{switch-case}$ would be deleted.

For example, the call-site of `uv__fs_work` in `uv_fs_access` would be replaced to `uv_fs_access.12` (12 is `ACCESS`'s enum value).`uv_fs_access.12` contains only the `access` case-branch and other case-branchs like `uv__fs_write_all` would be deleted. Therefore, this kind of false positives would be eliminated.

```
static void Read(const FunctionCallbackInfo<Value>& args) {
  AsyncCall(..., uv_fs_read, ...);
}
static void Unlink(const FunctionCallbackInfo<Value>& args) {
  AsyncCall(..., uv_fs_unlink, ...);
}
static void RMDir(const FunctionCallbackInfo<Value>& args) {
  AsyncCall(..., uv_fs_rmdir, ...);
}
FSReqBase* AsyncCall(..., Func fn, Args... fn_args) { AsyncCall.uv_fs_read
  [ ... ]
  return AsyncDestCall(..., fn, fn_args...);
}
FSReqBase* AsyncDestCall(..., Func fn, Args... fn_args) { AsyncDestCall.uv_fs_read
  [ ... ]
  req_wrap->Dispatch(fn, fn_args..., after);
}
int ReqWrap<T>::Dispatch(LibuvFunction fn, Args... args) { Dispatch.uv_fs_read
  [ ... ]
  CallLibuvFunction<T, LibuvFunction>::Call( fn, ...);
}
struct CallLibuvFunction<ReqT, int(*)(uv_loop_t*, ReqT*, Args...)> {
  [ ... ]
  static int Call(T fn, uv_loop_t* loop, ReqT* req, PassedArgs... args) {
    return fn(loop, req, args...);                             Call.uv_fs_read
  }
};
```

**Figure 6: Function-pointer parameter context-insensitive.**

*5.2.2 Function-Pointer Parameter Resolution.* We summarize the code patterns to locate functions for function-pointer parameter context-sensitive analysis: ❶ There exists more than one function pointer parameter in the function; ❷ The function pointer parameter would be called within the function. In Figure 6, `Call` satisfies these code patterns.

For each call-site of functions that match the pattern: ❶ HODOR performs inter-procedural and context-sensitive data-flow analysis for each call-site to trace the source of the function pointer object. ❷ If the depth of the call chain that locates the function pointer object is 1, $\langle$ $func_{caller}$, $func_{callee}$, $func_{fp}$ $\rangle$ are recorded. $func_{callee}$ is the function that contains function pointer parameters, and $func_{fp}$ is the function pointed by the function pointer parameter. We perform function clone on $func_{callee}$. The cloned function is named $func_{callee}.func_{fp}$, and the call-site in the cloned function is replaced to call the resolved function pointer. ❸ If the call chain depth is greater than 1, $\langle$ $func_{caller}$, $func_{callee}$, $*\rangle$ are recorded layer by layer backward until the function pointer object is determined to specific $func_{fp}$ point. Then we perform function clone to the functions of entire call chain from top to down. $func_{callee}$ will be cloned to $func_{callee}.func_{fp}$ in each layer, and the call-site in the cloned function is replaced to invoke the corresponding modified callee function except the last layer, which is replaced to call the resolved function pointer. For example, `AsyncCall` is cloned to three versions: `AsyncCall.uv_fs_read`, `AsyncCall.uv_fs_unlink` and `AsyncCall.uv_fs_rmdir`. The similar logic is performed in `AsyncDestCall`, `Dispatch` and `Call`. Eventually `uv_fs_read` is called through the chain: `Read` → `AsyncCall.uv_fs_read` → `AsyncDestCall.uv_fs_read` → `Dispatch.uv_fs_read` → `Call.uv_fs_read` and the `fn` in `Call.uv_fs_read` is resolved to `uv_fs_read`. Therefore, for each different value of the function pointer parameter, the call chain is different, which can eliminate the false positives.

We perform partial context-sensitive data-flow analysis to determine the assignment of the variables propagating to the branch selection variables and the source of the function pointer objects for the above two cases. The backward data-flow analysis starts with the branch selection variable or function pointer parameter, then it stops until the variable or the parameter is assigned with constants or actual function. When we achieve the endpoint of backward data-flow analysis, we perform a use-define chain analysis based on LLVM pass to determine whether the variable is assigned with one or more than one constant or actual function in the intra-procedural data flow analysis. When the variable is assigned with a single constant or actual function through different calling contexts, we perform context-sensitive analysis. In contrast, for the variable assigned with different constants or actual functions, we use context-insensitive analysis. As for the inter-procedural data-flow analysis, we additionally locate the starting point of the analysis at the callsite, record the context of different callsites, recursively perform intra-procedural analysis to complete inter-procedural data-flow analysis.

*5.2.3 Libc Analysis.* Almost all the dependency modules depend on `libc` library to interact with OS kernel. The popular `glibc` [52] library cannot be compiled into LLVM IR, thus we cannot utilize `SVF` to analyze it.

Previous works [28, 29] proposed methods for `glibc` analysis, but these analysis methods will introduce false positives and false negatives. `temporal- specialization` [28] utilizes GCC's RTL IR to analyze `glibc`. ❶ This method will lose some important system calls, such as `CLONE` system call in `pthread_create` and the `EXECVE` system call in `system`, etc.; ❷ It also connects some unrelated system calls to libc functions: `read` contains 41 system calls including `VFORK`, but actually the `read` function only contains one syscall `READ`. `Saphire` [29] utilizes `objdump` to generate the assembly of `libc.so` and analyzes the assemble text to collect system calls of libc functions. This method cannot recognize the `CLONE` system call in `pthread_create` and the `EXECVE` system call in `system` due to the lack of pointer analysis.

Musl libc [53] is a lightweight libc library compatible with Node.js and can be compiled by LLVM. We replaced `glibc` with `musl` to construct more precise call graphs. Musl libc uses the `inlineAsm` statement to set the register numbers including the syscall number and then call the `syscall` instruction. We implement an LLVM pass to conduct data-flow analysis to identify each `inlineAsm` statement related to `syscall` and figure out the syscall number. Musl libc also encapsulates `syscall` in C function. We compiled `Musl` with `-O3` optimization to inline these functions at their call-sites, so that all the encapsulated functions are expanded to `inlineAsm` statement and our data-flow analysis can gain the correct result.

## 5.3 Building Mappings

In this stage, we build API mappings to infer the system call whitelist between builtin modules, binding modules, and dependency modules. Algorithm 1 illustrates how HODOR generates mappings of the three layers.

**Algorithm 1:** Mapping generation

```
   Data: Call graphs of builtin modules cg.builtin, call graphs of binding modules and
         dependencies cg_bottom, call graphs of libc cg.libc, LLVM IR of binding
         modules ir_bind
   Result: Output mapping dict M
1  M.builtin ← {} ;                          /* Mappings of builtin modules */
2  forall cg.Module ∈ cg_builtin do
3  |  forall method ∈ module.exports do
4  |  |  M.builtin.module.method ← {};
5  |  |  Callers C invoked by the method by traversing cg.Module;
6  |  |  forall c ∈ C do
7  |  |  |  if c == internalBinding then
8  |  |  |  |  M.builtin.module.method ← (module, method);
9  M.binding ← {} ;                           /* Mappings of binding modules */
10 forall ir.Module ∈ ir_bind do
11 |  forall method_bind ∈ ir.Module do
12 |  |  M.binding.module.method ← func;
13 M.depend ← {} ;                            /* Mappings of dependencies */
14 forall func ∈ M.bindings.module.method do
15 |  M_depend.module.method ← {};
16 |  Callers C invoked by the function by traversing cg.bottom;
17 |  forall c ∈ C do
18 |  |  if c == libc then
19 |  |  |  M.depend.module.method ← libc;
20 forall libc ∈ M.depend.module.method do
21 |  M.depend.module.method ← {};
22 |  Callers C invoked by the function by traversing cg.libc;
23 |  forall c ∈ C do
24 |  |  if c ∈ syscall then
25 |  |  |  M.depend.module.method.libc ← syscall;
26 return M.builtin, M.binding, M.depend;
```

#### 5.3.1 Building Mappings for Builtin Modules.

Builtin modules utilize `module.export` or `export` object to export the builtin modules to Node.js applications and utilize `internalBinding` method to import the binding modules (i.e., binding method). To map the builtin methods and binding methods, we traverse the call graph (i.e., `cg.Module`) of each builtin module (Line 2-12). The entry point of the call graph is the method (i.e., `M.builtin.module.method`) list in the `module.export` object or `export` object. When the method invokes a method of an object which is imported by the `internalBinding` method, we record the method as the method of a binding module (Line 8). The name of the binding module is the argument of the `internalBinding`, and the name of the binding method is the property of the binding module.

#### 5.3.2 Building Mappings for Binding Modules.

As illustrated in Section 2.1, `node_module` structure is used to register binding modules to the builtin module layer. We develop a LLVM Pass to locate and analyze `node_module` structures (Line 13-18). We first traverse all global variables in the LLVM IR (i.e., `ir.Module`) to locate of lib node and locate the `node_module` structures that are used to register binding modules by matching the variable names and structure types. For each `node_module` structure, we record the binding module (i.e., `M.bindings.module`) and the `Initialize` function. As `Initialize` function utilizes set-method functions (including `SetMethod`, `SetMethodNoSideEffect`, etc [54]) to bind the binding methods (i.e., `M.bindings.module.method`) and C++ functions defined in binding layer, we traverse the `Initialize` function and extract set-method functions (Line 16). By analyzing the parameters of the call-sites of the set-method functions, we record mappings between binding methods and C++ functions. Finally, we collect the binding modules and the mapping between the builtin methods and the C++ functions. And the binding modules can be used on builtin module layer.

**Algorithm 2:** Whitelist generation

```
   Data: Call graph of Node.js Application cg_app, mapping sets M
   Result: Output whitelist W
1  wl.main ← {};
2  wl.pool ← {};
3  Callers C invoked application by traversing cg_app;
4  forall c ∈ C do
5  |  if c ∈ M.native.c then
6  |  |  forall b ∈ M.native.c do
7  |  |  |  if f ∈ M.bindings.b then
8  |  |  |  |  forall sys ∈ M.depend.f do
9  |  |  |  |  |  if b ∈ builtin_threadpool then
10 |  |  |  |  |  |  wl.pool ← sys;
11 |  |  |  |  |  else
12 |  |  |  |  |  |  wl.main ← sys;
13 return W
```

#### 5.3.3 Building Mappings for Dependency Modules.

We identify the system calls utilized by the C++ functions collected in the previous stage (Line 20-25). We first traverse the call graph of dependency modules, starting with the C++ functions that map binding methods. Then we collect `libc` functions invoked by the C++ functions at Line 22. For `libc` functions, we collect the system calls utilized by the `libc` functions by traversing the `libc` library call graph. Finally, we collect the mappings between the C++ function and system calls.

Moreover, we also identify the builtin methods that submit tasks to the thread pool, and the tasks they submit. As the main thread can submit tasks to the thread pool via `uv__work_submit` or `uv_queue_work` function, we first transverse the call graphs of Node.js framework to identify which methods use the thread pool. Furthermore, as `uv__work_submit` passes the function pointer in the fourth parameter to the thread pool for execution, we collect the tasks that are submitted to the thread pool by analyzing the call-site of `uv__work_submit`.

## 5.4 Whitelist Generation

Algorithm 2 illustrates how HODOR generate the whitelist the for Node.js application. We need to identify the method used by the applications. Since the application use `require` method to load builtin modules and dependent packages, we first traverse the call graph of the application, and identify `require` method call (Line 3). We identify the builtin methods and modules (i.e., *M.bindings.b*), and generate system call list for the application by linking builtin methods with the mappings constructed in the Building Mappings stage (on Line 4-9). Furthermore, to provide a fine-grained protection, we divide the system call list into system call list of main thread and system call list of thread pool. In Section 5.3, we collect the builtin methods that utilize the thread pool (`builtin_threadpool`). When generating the system call, if the builtin method is in the builtin thread pool (i.e., `builtin_threadpool` (Line 9), we add the system calls used by the tasks (i.e., dependency function `M.depend.f`) in thread-pool whitelist(Line 10), otherwise we add the system calls to the main thread whitelist (Line 12). Finally, we get the main thread whitelist and thread-pool whitelist.

## 5.5 HODOR Building

*5.5.1 Seccomp Implementation.* We propose a system-call level protection mechanism HODOR. As discussed in Section 2.1, Node.js applications can be divided into the applications that use the thread-pool thread and applications that do not use the thread-pool thread. For different types of applications, we adopt different loading mechanisms. For **thread pool required applications**, we first install the filter for the thread-pool thread based on the thread-pool whitelist and then install the filter for the main thread to prevent the thread pool thread from inheriting the main thread filter. Specifically, before the program is loaded, the thread pool will be executed first. We utilize `libseccomp` [55] library and add the code to the load filter after thread pool initialization. Then, before the main thread loads the applications, we load the main thread filter by using the `node-seccomp`. `node-seccomp` is a Node.js package that wraps around the libseccomp C library [56]. For **thread pool dis-required applications**, we only load the main thread filter. Moreover, HODOR collects system calls, which we refer to as engine-required system calls. These system calls are required by the pure Node.js engine without Node.js applications running on it. We collect these system calls by the dynamic tracing (`strace` tool [48]), and we add them to the whitelists.

*5.5.2 Read/write Permission Restrictions.* The state-of-art protection mechanisms based on `seccomp` do not restrict `read` and `write` system calls due to the requirements of application and process [21–29]. However, the lack of fine-grained restriction of `read` and `write` system calls allows the attackers to read or write sensitive files. Due to the wide use of `read` and `write` system calls in the Node.js engine (file operations or lower-level inter-thread communication), system call limitation mechanism can not shrink the attack surface related to these two system calls. In this work, we isolate the file system through `chroot` mechanism of `linux` to temporarily limit the root directory to the application directory when running the Node.js application. In this way, the attacker will not be able to modify the files outside the application directory for conducting corresponding attacks. Furthermore, we also switch the ownership of the files in the application directory to a high-privileged user and then set the files to read-only to prevent attackers from modifying the JavaScript code of the application. As the syscall for setting file permissions or switching the `uid` or `gid` are restricted by HODOR in most cases, the ability of attackers to tamper files can be further limited.

*5.5.3 Attack Surface Limitation.* Next, we will introduce the ability of HODOR to reduce the attack surface of several kinds of popular real-world ACE attacks referring to threat list of Node.js [57, 58]. ❶ **Gadget chain attacks** [13]. Prototype pollution attack is one of gadget chain attacks and is one of the most popular attacks of Node.js applications. The attacker gain the ability of ACE by polluting the arguments of `eval` or `child_process.exec` method. HODOR could mitigate gadget chain attacks in two aspects. If the application requires `child_process.exec` method, the attacker will not gain ACE ability as `EXECVE` system call is restricted. And if `exec` method or `eval` is allowed, HODOR protects the applications against the critical operations by system call limitation. ❷ **Injection attacks** [15]. Injection attacks include command injection, template injection, and code injection [58]. An attacker can achieve arbitrary code execution by injecting malicious code. By system call limitations, some malicious injection code cannot be executed. ❸ **Improper file access attacks** [59]. HODOR limits the file permissions that the application can access, and the attackers cannot read/write sensitive files. If the attacker can overwrite the permission allowed and executable files, the malicious code in the executable file cannot compromise the system further as HODOR applies system call limitation on the executable files. ❹ **Supply chain attacks** [11]. Supply chain attacks happen when the dependency chain exists vulnerabilities [11]. HODOR only provides the system calls used by the application. Therefore, although the attackers could exploit the code to obtain the ability of ACE through obsolete packages of the dependency chain, s/he cannot do any critical operations s/he wants due to the system calls limitations.

*5.5.4 Quantifying Attack Surface in Syscall Level.* We quantify the extent of syscall-related attack surface reduction. The base permissions of the application are the number of all system calls provided by the system.

$$S_{base} = |SYSCALL_{system}| \tag{1}$$

We generate whitelists for the main thread and thread pool and HODOR implements system call restriction on main thread the threads of thread pool:

$$S_{app} = |SYSCALL_{main\text{-}thread}| \cup |SYSCALL_{thread\text{-}pool}| \tag{2}$$

Finally we can quantify the extent of attack surface reduction:

$$SR = \frac{S_{app}}{S_{base}} \tag{3}$$

# 6 EVALUATION

We extensively evaluate HODOR by answering the following four research questions.

- **RQ1:** Can HODOR construct **sufficiently more precise call graphs** to achieve fine-grained system call level protection?
- **RQ2:** Can HODOR effectively **reduce the privileges of attackers** with arbitrary code/command execution attack ability?
- **RQ3:** How does HODOR compare with state-of-the-art tools?
- **RQ4:** What is the **runtime overhead** of the protection mechanism provided by HODOR?

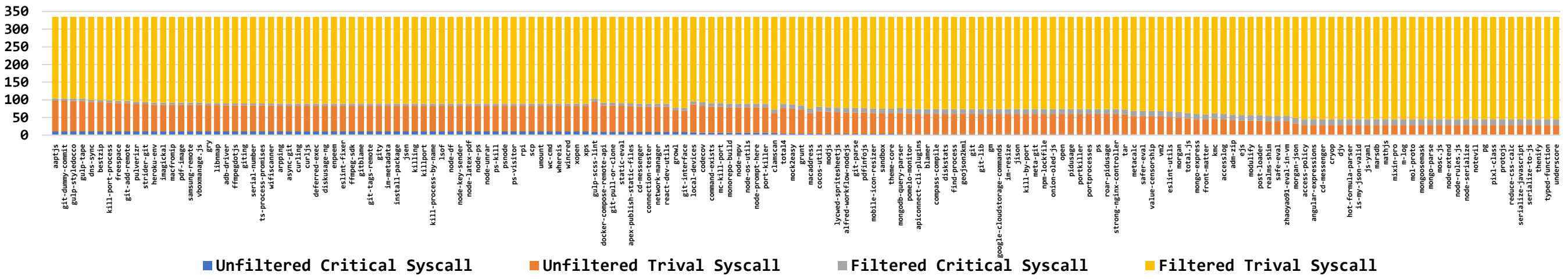

**Figure 7: System call for Node.js applications (RQ1).**

**Table 1: HODOR granularity at system call level and thread level (RQ1 ans RQ2).**

| Type | # of Package | Node.js w/ Musl | | Hodor | |
|---|---|---|---|---|---|
| | | # of CS | # of TS | # of CS | # of TS |
| Arb Comm Exec | 110 | 1,161 | 10,636 | 910 | 7,617 |
| Arb Code Exec | 58 | 243 | 3,352 | 110 | 2,294 |
| **Total** | 168 | 1,404 | 13,988 | 1,020 | 9,911 |

| Type | # of Package | # of MT | # of TP | # of MT | # of TP |
|---|---|---|---|---|---|
| Arb Comm Exec | 110 | 11,797 | 2,958 | 8,444 | 1,212 |
| Arb Code Exec | 58 | 3,595 | 1,218 | 2,362 | 496 |
| **Total** | 168 | 15,392 | 4,176 | 10,806 | 1,708 |

**CS**: Critical syscalls invocation; **TS**: Trivial syscalls invocation; **MT**: Main Thread system calls invocation; **TP**: Thread Pool system calls invocation;

Our experiments were conducted on a machine running 64-bit Ubuntu 18.04 with 64 AMD Ryzen Threadripper 3970X 28-Core Processor and 256GB RAM. The call graph building component for JavaScript code amounts to about 3.5K LoC and the call graph building component for C/C++ code amounts to about 2K LoC. The runtime enforcement component amounts to about 100 LoC.

**Result overview** We use Eq.3, i.e., *SR*, as a criterion to quantify the attack surface reduction ability. The smaller is the criterion, the better protection (attack surface reduction) can HODOR achieve. As shown in Figure 7 and Table 1, HODOR grants an average of 66 system calls each for the 168 packages in our evaluated dataset. For 58 packages with arbitrary code execution, among the 335 system calls provided by Linux kernel (i.e., $S_{base}$ is 335), the average *SR* is 12.37%. For 110 packages with arbitrary command execution, the average *SR* is 23.14%. on average, **HODOR can reduce the attack surface of Node.js applications to 19.42%.**

## 6.1 Dataset

We extensively collect those packages which are affected by arbitrary code execution or arbitrary command execution attacks from the GitHub Advisory Database [60] as follows. Firstly, we extracted the vulnerabilities with "arbitrary code execution" or "command execution" in the description. Then, we filtered out the vulnerabilities which are exploited from the client-side (e.g., Cross-Site Scripting attack). For the remaining vulnerabilities, we selected those readily available proof-of-concept exploits. In total, our experimental dataset consists of 169 packages. Among the 169 packages, HODOR crashes on one package. 58 packages suffer from code injection attacks , sandbox breakout attacks , file overwrite attacks , prototype pollution attacks , and template injection attacks . The left 111 packages suffer from arbitrary command execution attacks. The dataset fully covers the evaluated packages in MIR [20] and packages suffering from arbitrary code execution in `SecBench.js` [61]. Table 5 in Appendix shows the detailed descriptions of the vulnerable packages. We utilize this dataset as experiment targets for **RQ1, RQ2, RQ3** and **RQ4**. To evaluate the **RQ1**, we also take three large-scale real-world applications as extra datasets: `koa` [62], `express` [63] and `json-server` [64], whose github stars were no less than 30,000. Furthermore, to extensively evaluate **RQ4** (Runtime Overhead), we take two extra datasets: Node.js core tests [65] and 4 well-known web frameworks including `koa` [62], `fastify` [66], `express` [63], and `connect` [67], whose weekly downloads were no less than 500,000.

## 6.2 Call Graph Construction and Resulting Protection (RQ1)

We evaluate the improvement of HODOR in terms of call graphs construction and the resulting protection.

**Call Graphs Analysis** We first evaluate the achieved precision improvement of our call graphs construction optimizations, which is critical to accurately identify the system calls utilize by the Node.js application. In total, for 168 Node.js applications, ❶ Our static call graphs building method adds **3,555 edges** that are triggered by the execution of builtin methods of JavaScript and **547 edges** that are triggered by builtin methods of Node.js engine. In particular, 37 out of 168 packages (22.02%) invoke system calls that are triggered by the execution of builtin methods and would be undetectable without optimization. For example, `npm-lockfile` utilizes method `exec` of `child_process` module in the builtin objection `Promise`. A lack of analysis of them will lead to a too-restrictive whitelist, and the protection system will affect the normal operation of the application. ❷ As for dynamic call graph building methods, we build five test suites for the tested application on average and build combined static-dynamic call graphs for the application. In total, the combined static-dynamic call graphs building method adds and **1,271 nodes** and **2,474 edges**. These nodes and edges are generated due to the dynamic nature of JavaScript code, which cannot be captured by static call graphs analysis. ❸ For C/C++ code in the Node.js engine, HODOR totally cloned and replaced **36 functions** for switch-statement and **284 functions** for function-pointer parameter to implement partial context-aware analysis.

For the optimization of static call graphs construction for JavaScript code, we compared our method with the state-of-the-art JavaScript call graphs construction tools `JAM` [30] and `js-callgraph` [31]. Since we could not access the source code of `JAM`, we reproduce their comparison experiments and utilize the results they have done on six Node.js packages. We measured the precision (measured by comparing against dynamic call graphs building tool `NodeProf` [47]) according to `JAM` test method. As shown in Table 2, the result shows that HODOR finds that the coverage based on dynamically created edges is **92.39%**, compared to 79.77% for `JAM` and 52.49% for `js-callgraph`. The missing edges are triggered by the dynamic nature of JavaScript code. The results suggest that our optimizations for JavaScript code result in substantially higher coverage than the state-of-art JavaScript call graphs construction tools.

**Table 2: Experimental results for JavaScript call graph constructions (RQ1).**

| | HODOR | Jam | js-callgraph |
|---|---|---|---|
| **makeappicon 1.2.2** | 95.02% | 86.05% | / |
| **npm-git-snapshot 0.1.1** | 86.63% | 82.45% | 43.86% |
| **nodetree 0.0.3** | 83.33% | 70.65% | / |
| **jwtnoneify 1.0.1** | 93.43% | 71.43% | / |
| **npmgenerate 0.0.1** | 100% | 97.42% | 59.81% |
| **smrti 1.0.3** | 97.29% | 80.80% | 66.20% |
| **openbadges-issuer 0.4.0** | 91.05% | 75.85% | 40.08% |
| **Average** | 92.39% | 79.77% | 52.49% |

We choose 7 of 12 Node.js packages as `NodeProf` successfully builds call graph for these packages;
**Recall**: The coverage based on dynamically created edges;

For call graphs construction for C/C++ code, we compare our partial context-sensitive algorithm with state-of-art call graphs building tool `SVF`. Under the premise of running the program properly, we compared the size of the system call list generated by HODOR and `SVF`. In total, the size of the system call whitelist constructed by HODOR is **71.02%** (10,931/15,392) of the size of the whitelist constructed by `SVF`. In other words, **the optimization for C/C++ code significantly reduces the attack surface to 71.02% of `SVF`-built call graphs.** Furthermore, we divided the 335 system calls into 17 critical system calls (in Table 10 in Appendix) and 318 trivial system calls referring to `temporal-specialization` [28]. The critical system calls include command-execution-related system calls, network-related system calls, and permission-related system calls[3]. These critical system calls are not many but can cause enormous consequences such as creating scheduled tasks and reversing shells. As shown in Table 1, for critical system calls, our method could reduce the whitelist to **72.65%** (1,020/1,404) of whitelist generated by `SVF`-built call graphs. For trivial system calls, our method could reduce the whitelist to **70.85%** (9,911/13,988) of whitelist generated by `SVF`-built call graphs. More details are listed in Table 5 in Appendix. These results show that our optimization for C/C++ code help generate more precise system call lists for better shrinking the attack surface of ACE attacks.

**System Call Level Protection.** For 58 packages with arbitrary code execution, among the 335 system calls provided by Linux kernel (i.e., $S_{base}$ is 335), the average *SR* is 12.37For 110 packages with arbitrary command execution, the average *SR* is 23.14%. **HODOR can shrink the attack surface to 19.42% of the Node.js applications.**

We divided the system calls into engine-required system call type and app-required system call type. **Engine-required** system calls are a list of system calls required for the pure Node.js engine on which no application runs, and we record these system calls with the help of `strace` [48] utility. The engine-required system call set includes 28 system calls for the main thread and 15 system calls for the thread pool. The full list of the engine-required system calls is shown in Table 9 in Appendix. Other system calls are considered as **app-required** system calls, which are only used when the application calls the methods of builtin modules. Among the 168 packages, we observed 129 of 168 (76.79%) packages use builtin modules and methods. Therefore, in addition to the engine-required system call, HODOR also needs to use the mappings between builtin modules and system calls to infer the app-required system calls. 39 of 168 (23.21%) packages do not utilize the builtin modules, that is, do not need app-required system calls. All of them are packages that are compromised by arbitrary code attacks. Since no additional builtin methods are used, HODOR can reduce the attack surface to the maximum extent, i.e., the attacker could only use 28 system calls.

**Thread Level Protection.** HODOR adds filters on both the main threads and the threads of thread pool. As shown in Table 1, for the main thread, the attack surface can be reduced to 19.20% by HODOR on average. Among the 129 builtin-method-needed packages, we observe 66 packages (51.16%) invoke the system calls that are executed in the thread pool. For these packages, HODOR also adds filters on the threads of the thread pool, and the attack surface thread pool can be reduced to 7.73% by HODOR on average. For example, HODOR can identify `npm-lockfile` package utilize method `readFile`, `writeFile`, `stat`, and `mkdir` belongs to the builtin module `fs`, which are accomplished by the thread pool. Armed with HODOR, attackers cannot perform thread pool related exploits apart from `readFile`, `writeFile`, `stat`, and `mkdir`.

**Dynamic Command Execution** We extract and analyze the command arguments of command execution methods (i.e., `exec` and `spawn` methods of `child_process` module) and add the system calls required by these commands into the system call list of the main thread. For 98 packages, we add 4,887 system calls to the system calls, including 61 critical system calls and 4,826 trivial system calls.

**Code Coverage and HODOR's Effectiveness.** We reported the coverage of each of the evaluated benchmark in the test suite in the last three columns of Table 5 in the Appendix to show its effect on HODOR's effectiveness. The key observation here is that there is little correlation between the code coverage and the number of system calls in the whitelist of HODOR, i.e., the Pearson product-moment correlation coefficient between the package module covered line number and the whitelist size is -0.038, coefficient between the code coverage of package and the whitelist size is -0.027, and coefficient between the code coverage of package module and the whitelist size is -0.066 (more details in Figure 9 in the Appendix). The implication is that HODOR is able to precisely infer the system calls required regardless of the code coverage, which thus alone does not have a significant effect on HODOR's effectiveness.

**HODOR's Effect on Benign Applications.** We used the above test suites and three more real-world applications to evaluate HODOR's effect on the application's normal operation. Among the 169 packages, 168 packages run correctly, while HODOR crashed on the package `extra-asciinema` [68]. Note that as shown in Table 5 in Appendix, the average covered line of the package module is 69.79%, and the average cover line of the package including the dependent

[3]Note that `read`, `write`, and `unlink` are carefully addressed by limiting file permissions by HODOR as discussed in Section 5.5.2

module is 33.98%. The crashed package contains Python scripts, and there was an interaction between Python3 and JavaScript scripts at runtime. HODOR does not support to analyze Python scripts leading to the syscall list missing required by the Python scripts at runtime. Therefore, the package suffered from a too-restrictive list of system calls, causing it to crash with the protection of HODOR. For real-world applications testing, we deployed and accessed the websites. As shown in Table 3, the average covered line of the package module is 56.11%, and the average cover line of the package is 11.15%. The result shows that HODOR can be integrated into well-known applications and does not affect the normal functionalities.

**Takeaway:** The optimization of JavaScript call graph construction helps identify hidden required system calls for 23.21% packages. And the optimization in C/C++ level further reduces the system call permissions by 71.02% on average. They together enable HODOR to reduce the attack surface for the main thread to 19.20%, and reduce the attack surface of the thread pool thread to 7.73%, while not affecting the application's normal operation.

**Table 3: Experimental results for correctness (RQ1).**

| | Stars | LoC | # Syscall | % CL-1 | % CL-2 | Pass Ratio |
|---|---|---|---|---|---|---|
| **json-server** | 66.5K | 258,094 | 61 | 53.36% | 3.34% | 32/32 |
| **express** | 60.6K | 58,997 | 63 | 66.05% | 26.12% | 75/75 |
| **koa** | 33.9K | 291,226 | 64 | 48.92% | 3.98% | 28/28 |
| **Average** | **53.7K** | **202,772** | **62.67** | **56.11%** | **11.15%** | **135/135** |

**CL-1**: Covered line of the package module; **CL-2**: Covered line of the package;

## 6.3 Exploit Mitigation (RQ2)

We further evaluate the effectiveness of HODOR from the exploition perspective. Traditional runtime protection work usually uses the attack payload provided by the proof of concept when evaluating their effectiveness. However, the payload provided by proof of concept is far from enough, as the operations performed by attackers in reality can be sophisticated and require system calls, while the proof of concept is generally simple, such as `print 123` or execute `ls`. Therefore, to extensively verify the effectiveness of our tool, **we construct different advanced attack payloads to simulate various dangerous behaviors of attackers, where a variety of critical system calls can be invoked**. Specifically, according to different vulnerability types, we construct seven payloads written in bash language for arbitrary code execution and seven payloads written in JavaScript language for arbitrary command execution. The details are shown in Table 8 in Appendix.

For the packages with arbitrary command execution, HODOR could successfully defend against 62.08% of the critical exploits and defend against 99.09% of real-world exploits. HODOR cannot block the execution of functions associated with arbitrary command execution (e.g., `child_process.exec`), but it can reduce the ability of the code that can be executed. For the packages with arbitrary code execution, HODOR could defend against all of the attacks with arbitrary code execution. On average, HODOR could effectively mitigate the execution of 91.63% critical exploits and defend against 51.72% of real-world exploits.

We demonstrate two case studies to further show the exploit mitigation capability of HODOR. ❶ For arbitrary command execution vulnerabilities, attackers can inject shell code to achieve arbitrary code execution. This type of vulnerability utilizes the `exec` method or `spawn` method of the `child_process` module. Such vulnerabilities utilize `exec` system calls, while due to the inheritance feature of `seccomp`, the system calls that `exec` can invoke are restricted. Package `dns-sync` use `child_process.execSync` method to execute the bash command. By command injections, the attackers could inject the bash code into the applications. HODOR limits the attack surface by applying the system call restrictions on the commands. Thus although the attacker can inject arbitrary command into the `execSync` method of `child_process` module, the critical exploits cannot execute successfully as HODOR disables the system calls required by critical exploits. ❷ For arbitrary code execution vulnerabilities, the attacker can inject JavaScript code to achieve the purpose of arbitrary code execution. Package `access-policy` does not use builtin methods, and the vulnerabilities are caused by the parameter injection of the `eval` method. HODOR provide `access-policy` with minimum permissions. Therefore although the attacker can inject arbitrary JavaScript code into the `eval` method, the exploits cannot execute successfully as HODOR disables the system calls required by critical exploits.

**Takeaway:** HODOR could effectively mitigate the execution of 73.59% exploits.

## 6.4 Comparison with Other Techniques (RQ3)

**Table 4: Comparison with state-of-art protection mechanisms.**

| Property | Mininode [18] | Mir [20] | HODOR |
|---|---|---|---|
| **Security mechanism** | Code debloating | Privilege reduction | Syscall limitation |
| **Runtime protection** | / | JavaScript level | Kernel level |
| **Source code change** | Debloating | Function Closure | Non-change |
| **Protection granularity** | / | Module | Application |

**Qualitative Comparison.** As shown in Table 4, we perform a qualitative comparison between HODOR and the state-of-art works [18, 20] for attack surface shrinking from four aspects. ***Security Mechanism.*** `Mininode` removes the unused code and `MIR` designs privilege reduction through context wrappers for the application, which both targets the JavaScript source code level (and may bring noticeable overhead). HODOR takes a complementary perspective to target the system call level for shrinking the attack surface. Mininode does not provide runtime protection. For attackers with ACE attack ability, s/he can still perform relevant critical operations. `MIR` provides runtime protections mainly at the JavaScript application level while HODOR provides runtime protection at the Kernel level via Syscall limitation. ***Source Code change.*** `Mininode` reduces the packages and `MIR` wraps the module with a function closure, which both need source code modification. Nevertheless, considering the dynamic nature of JavaScript, the source code level modifications might invalidate the protection mechanism and even disable the application to run properly. ***Protection Granularity.*** `MIR` is module-level protection where different module has different privilege. `Mininode` and HODOR are application-level protection.

**Quantitative Comparison.** We compare HODOR with state-of-art attack surface shrinking tool for Node.js applications, i.e., MIR, a permission inference and restriction system, which restricts the arbitrary code attack surface by adding a read-write-execute (RWX) permission model. MIR has two configurations, MIR-sa and MIR-da. MIR-sa generate default permissions for the package and MIR-da enforce the protection on the Node.js applications at runtime.

We utilize MIR-sa and MIR-da to run the 58 packages that suffer from arbitrary code attacks and check the permissions against the payload we constructed in Section 6.3. We manually audit the default permission generated by MIR-sa and run the tested application with MIR-da. MIR-sa failed to analyze 15 of 58 (25.86%) packages, as it can not generate proper permissions for these packages. Therefore, MIR-sa loads the most strict filter for them, which blocks their proper execution. MIR-da failed to run 17 of 58 (29.31%) packages. Among them, three failures were caused by program errors of MIR-da. 14 failures were caused by changes of MIR to the global variables of the applications. For the rest 41 packages, 99.39% of the critical exploits could be blocked under the permission models generated by MIR-sa. MIR-da introduced an overhead of 6.98%. Since MIR-da checks the permissions of the module every time methods or global variables are invoked. while the overhead of HODOR is 0.72% (more details in Sec. 6.5).

**Takeaway:** Compared to MIR and mininode, HODOR does not require source code modification and is capable of defending against a wider spectrum of attacks (additionally covering arbitrary command execution) with less runtime overhead.

## 6.5 Runtime Overhead (RQ4)

To evaluate the overhead introduced by libc library changes and HODOR additions, we ran these tests using Node.js in three states: Node.js with glibc, Node.js with Musl libc, and with HODOR Node.js (i.e. protected Node.js with Musl libc). Since Node.js is developed based on glibc libc, we standardize the results on when Node.js with glibc runs code.

For the Node.js core tests, the test case will run a statement containing a builtin method execution multiple times (200 in our experiments) and record the running time. Figure 8a shows that the total time of running the tests. The results show that the overhead introduced by libc library replacement is 1.27%, and the protection mechanism introduces 0.61% overhead further.

For the web frameworks of koa, fastify, express, and connect, we measure the overhead of HODOR by observing the response of the index page provided by the the example code in the framework tutorial. We conducted 2000 requests of index page visits and record the total response time of the server. As shown in Figure 8b, the replacement of libc library introduces 0.76% overhead on average and the protection mechanism introduces 2.80% overhead.

Figure 8c shows the runtime overhead of different attack type of the 168 packages. We find that libc replacement introduces 1.90% overhead on average, and the protection mechanism introduces 0.39% on average. Not surprisingly, we find packages that use fewest builtin methods (i.e., the packages effected by arbitrary code execution attack) introduce minimal overhead (0.22%), while packages that use more system call related operations (i.e., the packages effected by arbitrary command execution attack) introduce higher overhead (0.42%).

**Takeaway:** On average, the runtime overhead of HODOR is 0.61% for Node.js core tests, 2.80% for the web framework, and 0.39% for all the 168 packages, which is in general acceptable.

# 7 RELATED WORK

**System Call Limitation:** Many studies reduced the attack surface by limiting the system call set that the attacker can invoke in other application domains including software [25, 43], android applications [69], container applications [24, 28], linux applications [27] and PKU-based memory isolation systems [70], etc. For instance, the closest and recent work to HODOR proposed by Wyss *et al.* [43] introduces Latch for mediating the install-time capabilities of npm packages. It generates a system call manifest of install script and enforces it to prevent undesirable install-time behavior. Bulekov *et al.* [29] proposed an automatic approach for generating and applying system-call limitations to interpreted PHP applications. They performed static and dynamic analysis to build call graphs over the interpreter for the binary of the PHP application and interpreter. Compared with HODOR, Saphire scanned the AST to identify all built-in function calls, built call graphs over the compiled binary from Debian repositories, and applied protection to the entire process, which are coarse-grained for Node.js applications. Moreover, in addition to including the system calls used by the application, the whitelist also preserved the system calls that Node.js invokes at runtime. Ghavamnia *et al.* [28] was tailored for server applications. They presented an approach that could restrict the system call set further based on the execution phase. Ghavamnia *et al.* [24] proposed a solution for automatically generating limited system call policies for Docker containers including server application containers. Their work is orthogonal as Node.js provides capabilities to create a web server. Nevertheless, the system calls invoked by Node.js are related to the application. What's more, Node.js is a multi-threaded application, which needs thread granularity filtered limitation.

**Code Debloating, Privilege Reduction, and Security Isolation:** In this line of research, some studies removed the unused API or unused code of the application. Koishybayev *et al.* [18] leveraged static analysis to remove the unused code and dependencies of Node.js applications. Azad *et al.* [71] obtained the dead code of PHP applications by using the by dynamic analysis. Snyder *et al.* [72] evaluated the Web API of modern browsers and proposed a Web-API access extension for client-side users. Qian *et al.* [73] utilized a hybrid approach to determine the bloated units of Chromium and removed them. In the comparison of our work, these technologies could prevent attackers from further exploiting malicious code. However, as mentioned in Section 6.4, these techniques cannot limit arbitrary code execution attacks. Others implement privilege reduction. Bittau *et al.* [74] presented the system Wedge to splitting the application into fine-grained, least-privilege compartments. Gudka *et al.* [75] proposed the Security-Oriented Analysis of Application Programs (SOAAP) that could create isolated components for complex applications to limit privileges leaked. Vasilakis *et al.* [20] focused on the dynamic compromise on Node.js. They proposed a read-write-execute (RWX) permission module MIR for the Node.js application. MIR is close and the most recent to our work, however, as shown in section 6.4, this work cannot reduce the attack surface of some arbitrary code execution attacks. Others implement security isolation mechanism. For instance, Vasilakis *et al.* [76] introduced `BreakApp` that transforms a module into its own compartment and enforces security policies on module communication automatically based on the users' configuration. While the policies are generated manually. Ahmadpanah *et al.* [77] performed security analysis on Node-RED applications and proposed sandbox system `SandTrap` for the applications. `SandTrap` can enforce fine-grained access control policies on third-party applications.

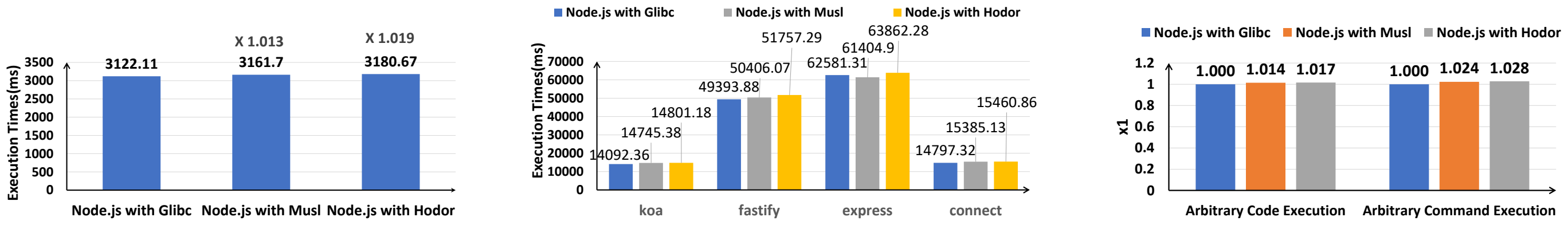

(a) Node.js Core Tests (b) Web Framework (c) Node.js Packages

**Figure 8: Runtime overhead of Node.js core tests, web framework and applications under the protection of HODOR (RQ4).**

**JavaScript Program Analysis:** In recent years, researchers have developed static analyzers (such as JSAI [78], TAJS [79], WALA [80], and SAFE [81]) and dynamic analysis (such as Jalangi [82] and Nodeprof [47]) to understand behaviors of JavaScript programs and to detect their bugs in a sound manner. Analysis precision and performance are being increased. Andreasen *et al.* [83] presented a static dataflow analysis for JavaScript with high degrees of context sensitivity. Park *et al.* [84] presented Loop-Sensitive Analysis (LSA) approach to enhance the static analysis precision in loops. Stein *et al.* [85] presented a conventional non-relational static dataflow analysis with a value refinement mechanism to increase precision on demand at critical locations. Park *et al.* [86] proposed dynamic shortcuts to switch between abstract and concrete execution during JavaScript static analysis in a sound way. By employing more efficient static and dynamic analysis, we can generate more fine-grained system call whitelist for the Node.js applications.

**Node.js Application Analysis:** The past two years have seen many researches on Node.js security. Nielsen *et al.* [30] proposed an approach to build call graphs for Node.js applications, which can be used for security scanning. Li and Kang *et al.* [58] leveraged a static analysis to generate a graph structures, called Object Dependence Graph (ODG), for detecting Node.js vulnerabilities. staicu *et al.* [87] presented an automatically extracting taint specifications for Node.js applications. Xiao *et al.* [42] found the hidden property abusing (HPA) attack, and designed a tool to detect and verify the vulnerability. Compared to the previous work focusing on mining Node.js application vulnerabilities, our work aims to provide runtime protection for Node.js applications.

# 8 DISCUSSION & CONCLUSION

We now conclude by considering potential limitations of HODOR as well as opportunities for future work: ❶ HODOR does not analyze native extensions [88] of Node.js applications yet, as the evaluated applications do not use them. They will be considered for future through LLVM pass extensions (in Section 5.3.2). ❷ HODOR provides application-level enforcement mechanisms that is more coarse-grained than module-level mechanisms (e.g., `mir` [20]). Moreover, HODOR cannot reduce the attack surface to zero due to system call requirements of pure Node.js engine. In the future, it can be combined with other protection mechanisms at different levels to complement each other. ❸ As now, our static analysis for Node.js application only supports the module system CommonJS, which lacks portability on ES6 and AMD module systems. In the future, we will consider supporting ES6 and AMD module systems.

In this paper, we propose HODOR, a novel permission restriction system in the lower system call level for Node.js applications to shrink the attack surface. HODOR first generates fine-grained system call permissions for the Node.js applications, benefiting from multiple significant optimizations to improve state-of-the-art methods in call graph construction for both JavaScript code and C/C++ code. A more precise CFG can also benefit a broad range of applications including supply chain attack detection, program security analysis, code navigation and refactoring, etc. Then, based on the identified required permissions, HODOR creates the whitelists and implements the runtime restriction in thread granularity level with the seccomp mechanism. Armed with HODOR, we apply HODOR on 168 real-world Node.js applications suffering from arbitrary code/-command execution attacks. Extensive experiments show that the attack surfaces of the vulnerable packages reduce to 19.42%, while introducing negligible runtime overhead, i.e., less than 3%.

# ACKNOWLEDGMENTS

The authors would like to thank the reviewers for their valuable feedback during the revision process. This work was supported by the National Key Research and Development Program of China (No. 2021YFB3101402).

# REFERENCES

[1] Node.js. Node.js. https://nodejs.org/. Accessed: 2023.

[2] PayPal Tech Blog Admin Account. Paypal Engineering. Node.js at PayPal. https://medium.com/paypal-engineering/node-js-at-paypal-4e2d1d08ce4f.

[3] LinkedIn Moved From Rails To Node: 27 Servers Cut And Up To 20x Faster. http://highscalability.com/blog/2012/10/4/linkedin-moved-from-rails-to-node-27-servers-cut-and-up-to-2.html.

[4] Matthew Baxter-Reynolds. Here's why you should be happy that microsoft is embracing node.js. https://www.theguardian.com/technology/blog/2011/nov/09/programming-microsoft.

[5] Netflix TechBlog. Nodejs. https://netflixtechblog.com/tagged/nodejs.

[6] npm. https://www.npmjs.com/.

[7] libuv. Asynchronous I/O made simple. https://libuv.org/. Accessed: 2023.

[8] V8. V8 JavaScript engine. https://v8.dev/. Accessed: 2023.

[9] Markus Zimmermann, Cristian-Alexandru Staicu, Cam Tenny, and Michael Pradel. Small world with high risks: A study of security threats in the npm ecosystem. In *28th USENIX Security Symposium (USENIX Security 19)*, pages 995–1010, 2019.

[10] Chengwei Liu, Sen Chen, Lingling Fan, Bihuan Chen, Yang Liu, and Xin Peng. Demystifying the vulnerability propagation and its evolution via dependency trees in the npm ecosystem. *arXiv preprint arXiv:2201.03981*, 2022.

[11] Nusrat Zahan, Thomas Zimmermann, Patrice Godefroid, Brendan Murphy, Chandra Maddila, and Laurie Williams. What are weak links in the npm supply chain? In *2022 IEEE/ACM 44th International Conference on Software Engineering: Software Engineering in Practice (ICSE-SEIP)*, pages 331–340. IEEE, 2022.

[12] Bodin Chinthanet, Raula Gaikovina Kula, Shane McIntosh, Takashi Ishio, Akinori Ihara, and Kenichi Matsumoto. Lags in the release, adoption, and propagation of npm vulnerability fixes. *Empirical Software Engineering*, 26(3):1–28, 2021.

[13] Mikhail Shcherbakov, Musard Balliu, and Cristian-Alexandru Staicu. Silent spring: Prototype pollution leads to remote code execution in node. js. In *USENIX Security Symposium 2023*, 2023.

[14] Song Li, Mingqing Kang, Jianwei Hou, and Yinzhi Cao. Detecting node. js prototype pollution vulnerabilities via object lookup analysis. In *Proceedings of the 29th ACM Joint Meeting on European Software Engineering Conference and Symposium on the Foundations of Software Engineering*, pages 268–279, 2021.

[15] Cristian-Alexandru Staicu, Michael Pradel, and Benjamin Livshits. Synode: Understanding and automatically preventing injection attacks on node. js. In *NDSS*, 2018.

[16] Wikipedia. Arbitrary code execution. https://en.wikipedia.org/wiki/Arbitrary_code_execution.

[17] Prototype Pollution. https://security.snyk.io/vuln/SNYK-JS-LODASHMERGE-173732.

[18] Igibek Koishybayev and Alexandros Kapravelos. Mininode: Reducing the attack surface of node. js applications. In *23rd International Symposium on Research in Attacks, Intrusions and Defenses (RAID 2020)*, pages 121–134, 2020.

[19] Alexi Turcotte, Ellen Arteca, Ashish Mishra, Saba Alimadadi, and Frank Tip. Stubbifier: debloating dynamic server-side javascript applications. *Empirical Software Engineering*, 27(7):1–36, 2022.

[20] Nikos Vasilakis, Cristian-Alexandru Staicu, Grigoris Ntousakis, Konstantinos Kallas, Ben Karel, André DeHon, and Michael Pradel. Preventing dynamic library compromise on node. js via rwx-based privilege reduction. In *Proceedings of the 2021 ACM SIGSAC Conference on Computer and Communications Security*, pages 1821–1838, 2021.

[21] Sandboxing ChromeOS system services. https://chromium.googlesource.com/chromiumos/docs/+/HEAD/sandboxing.md.

[22] Seccomp security profiles for Docker. https://docs.docker.com/engine/security/seccomp/.

[23] Nuno Lopes, Rolando Martins, Manuel Eduardo Correia, Sérgio Serrano, and Francisco Nunes. Container hardening through automated seccomp profiling. In *Proceedings of the 2020 6th International Workshop on Container Technologies and Container Clouds*, pages 31–36, 2020.

[24] Seyedhamed Ghavamnia, Tapti Palit, Azzedine Benameur, and Michalis Polychronakis. Confine: Automated system call policy generation for container attack surface reduction. In *23rd International Symposium on Research in Attacks, Intrusions and Defenses (RAID 2020)*, pages 443–458, 2020.

[25] Nicholas DeMarinis, Kent Williams-King, Di Jin, Rodrigo Fonseca, and Vasileios P Kemerlis. Sysfilter: Automated system call filtering for commodity software. In *23rd International Symposium on Research in Attacks, Intrusions and Defenses (RAID 2020)*, pages 459–474, 2020.

[26] Daoyuan Wu, Debin Gao, Yingjiu Li, and Robert H Deng. Seccomp: Towards practically defending against component hijacking in android applications. *arXiv preprint arXiv:1609.03322*, 2016.

[27] Claudio Canella, Mario Werner, Daniel Gruss, and Michael Schwarz. Automating seccomp filter generation for linux applications. In *Proceedings of the 2021 on Cloud Computing Security Workshop*, pages 139–151, 2021.

[28] Seyedhamed Ghavamnia, Tapti Palit, Shachee Mishra, and Michalis Polychronakis. Temporal system call specialization for attack surface reduction. In *29th USENIX Security Symposium (USENIX Security 20)*, pages 1749–1766, 2020.

[29] Alexander Bulekov, Rasoul Jahanshahi, and Manuel Egele. Saphire: Sandboxing php applications with tailored system call allowlists. In *30th USENIX Security Symposium (USENIX Security 21)*, pages 2881–2898, 2021.

[30] Benjamin Barslev Nielsen, Martin Toldam Torp, and Anders Møller. Modular call graph construction for security scanning of node. js applications. In *Proceedings of the 30th ACM SIGSOFT International Symposium on Software Testing and Analysis*, pages 29–41, 2021.

[31] Asger Feldthaus, Max Schäfer, Manu Sridharan, Julian Dolby, and Frank Tip. Efficient construction of approximate call graphs for javascript ide services. In *2013 35th International Conference on Software Engineering (ICSE)*, pages 752–761. IEEE, 2013.

[32] Yulei Sui and Jingling Xue. Svf: interprocedural static value-flow analysis in llvm. In *Proceedings of the 25th international conference on compiler construction*, pages 265–266, 2016.

[33] dns-sync. https://www.npmjs.com/package/dns-sync.

[34] Node.js. Nodejs document. https://nodejs.org/api.

[35] The Linux Kernel. Seccomp BPF (SECure COMPuting with filters). https://www.kernel.org/doc/html/v4.16/userspace-api/seccomp_filter.html.

[36] Berkeley Packet Filter. https://en.wikipedia.org/wiki/Berkeley_Packet_Filter.

[37] Growl for nodejs. https://www.npmjs.com/package/growl.

[38] Gabriel Ferreira, Limin Jia, Joshua Sunshine, and Christian Kästner. Containing malicious package updates in npm with a lightweight permission system. In *2021 IEEE/ACM 43rd International Conference on Software Engineering (ICSE)*, pages 1334–1346. IEEE, 2021.

[39] André Takeshi Endo and Anders Møller. Noderacer: Event race detection for node. js applications. In *2020 IEEE 13th International Conference on Software Testing, Validation and Verification (ICST)*, pages 120–130. IEEE, 2020.

[40] James C Davis, Eric R Williamson, and Dongyoon Lee. A sense of time for Javascript and node. js: First-class timeouts as a cure for event handler poisoning In *27th {USENIX} Security Symposium ({USENIX} Security 18)*, pages 343–359, 2018.

[41] Cristian-Alexandru Staicu and Michael Pradel. Freezing the Web: A Study of ReDoS Vulnerabilities in JavaScript-based Web Servers. In *USENIX Security Symposium*, pages 361–376, 2018.

[42] Feng Xiao, Jianwei Huang, Yichang Xiong, Guangliang Yang, Hong Hu, Guofei Gu, and Wenke Lee. Abusing hidden properties to attack the node.js ecosystem. In *30th USENIX Security Symposium (USENIX Security 21)*, pages 2951–2968, 2021.

[43] Elizabeth Wyss, Alexander Wittman, Drew Davidson, and Lorenzo De Carli. Wolf at the door: Preventing install-time attacks in npm with latch. In *Proceedings of the 2022 ACM on Asia Conference on Computer and Communications Security*, pages 1139–1153, 2022.

[44] Zhihao Bai, Ke Wang, Hang Zhu, Yinzhi Cao, and Xin Jin. Runtime recovery of web applications under zero-day redos attacks. In *2021 IEEE Symposium on Security and Privacy (SP)*, pages 1575–1588. IEEE, 2021.

[45] Standard built-in objects. https://developer.mozilla.org/en-US/docs/Web/JavaScript/Reference/Global_Objects.

[46] Saba Alimadadi, Di Zhong, Magnus Madsen, and Frank Tip. Finding broken promises in asynchronous javascript programs. *Proceedings of the ACM on Programming Languages*, 2(OOPSLA):1–26, 2018.

[47] NodeProf on GraalVM - Panathon 2018. https://github.com/Haiyang-Sun/nodeprof.js.

[48] Strace Utility. https://strace.io/.

[49] fail0verflow. Clang. https://clang.llvm.org.

[50] Ian A Mason. Whole Program LLVM. https://github.com/travitch/whole-program-llvm.

[51] llvm-link. https://llvm.org/docs/CommandGuide/llvm-link.html.

[52] The GNU C Library (glibc). https://www.gnu.org/software/libc/.

[53] musl libc. https://www.musl-libc.org.

[54] node_buffer.cc. https://github.com/nodejs/node/blob/main/src/node_buffer.cc.

[55] libseccomp. https://github.com/seccomp/libseccomp.

[56] node-seccomp. https://www.npmjs.com/package/node-seccomp.

[57] Node.js. Node.js Security Best Practices. https://nodejs.org/en/docs/guides/security/. Accessed: 2022.

[58] Mining node.js vulnerabilities via object dependence graph and query. In *31st USENIX Security Symposium (USENIX Security 22)*, Boston, MA, August 2022. USENIX Association.

[59] Node.js Vulnerability Cheatsheet. https://www.shiftleft.io/blog/node.js-vulnerability-cheatsheet/. Accessed: 2022.

[60] GitHub Advisory Database. https://github.com/advisories.

[61] Masudul Hasan Masud Bhuiyan, Adithya Srinivas Parthasarathy, Nikos Vasilakis, Michael Pradel, and Cristian-Alexandru Staicu. Secbench. js: An executable security benchmark suite for server-side javascript. In *International Conference on Software Engineering (ICSE)*, 2023.

[62] koa. https://www.npmjs.com/package/koa.

[63] express. https://www.npmjs.com/package/express.

[64] JSON Server. https://github.com/typicode/json-server.

[65] Node.js Core Tests. https://github.com/nodejs/node/tree/main/test.

[66] fastify. https://www.npmjs.com/package/fastify.

[67] connect. https://www.npmjs.com/package/connect.

[68] wolfram77. extra-asciinema. https://www.npmjs.com/package/extra-asciinema.

[69] Yingjiao Niu, Yuewu Wang, Shijie Jia, Quan Zhou, Lingguang Lei, Qionglu Zhang, and Xinyi Zhao. Enhancing the security of mobile device management with seccomp. In *Journal of Physics: Conference Series*, volume 1646, page 012138. IOP Publishing, 2020.

[70] David Schrammel, Samuel Weiser, Richard Sadek, and Stefan Mangard. Jenny: Securing syscalls for pku-based memory isolation systems. In *Proceedings of the 31th USENIX Security Symposium*, 2022.

[71] Babak Amin Azad, Pierre Laperdrix, and Nick Nikiforakis. Less is more: quantifying the security benefits of debloating web applications. In *28th USENIX*

*Security Symposium (USENIX Security 19)*, pages 1697–1714, 2019.

[72] Peter Snyder, Cynthia Taylor, and Chris Kanich. Most websites don't need to vibrate: A cost-benefit approach to improving browser security. In *Proceedings of the 2017 ACM SIGSAC Conference on Computer and Communications Security*, pages 179–194, 2017.

[73] Chenxiong Qian, Hyungjoon Koo, ChangSeok Oh, Taesoo Kim, and Wenke Lee. Slimium: Debloating the chromium browser with feature subsetting. In *Proceedings of the 2020 ACM SIGSAC Conference on Computer and Communications Security*, pages 461–476, 2020.

[74] Andrea Bittau, Petr Marchenko, Mark Handley, and Brad Karp. Wedge: Splitting applications into reduced-privilege compartments. USENIX Association, 2008.

[75] Khilan Gudka, Robert NM Watson, Jonathan Anderson, David Chisnall, Brooks Davis, Ben Laurie, Ilias Marinos, Peter G Neumann, and Alex Richardson. Clean application compartmentalization with soaap. In *Proceedings of the 22nd ACM SIGSAC Conference on Computer and Communications Security*, pages 1016–1031, 2015.

[76] Nikos Vasilakis, Ben Karel, Nick Roessler, Nathan Dautenhahn, André DeHon, and Jonathan M Smith. Breakapp: Automated, flexible application compartmentalization. In *NDSS*, 2018.

[77] Mohammad M Ahmadpanah, Daniel Hedin, Musard Balliu, Lars Eric Olsson, and Andrei Sabelfeld. Sandtrap: Securing javascript-driven trigger-action platforms. In *USENIX Security Symposium (USENIX Security 2021)*, 2021.

[78] Vineeth Kashyap, Kyle Dewey, Ethan A Kuefner, John Wagner, Kevin Gibbons, John Sarracino, Ben Wiedermann, and Ben Hardekopf. Jsai: A static analysis platform for javascript. In *Proceedings of the 22nd ACM SIGSOFT international symposium on Foundations of Software Engineering*, pages 121–132, 2014.

[79] Simon Holm Jensen, Anders Møller, and Peter Thiemann. Type analysis for javascript. In *International Static Analysis Symposium*, pages 238–255. Springer, 2009.

[80] Stephen Fink and Stephen Fink. WALA–The TJ Watson Libraries for Analysis. https://github.com/wala/WALA. Accessed:2012.

[81] Hongki Lee, Sooncheol Won, Joonho Jin, Junhee Cho, and Sukyoung Ryu. Safe: Formal specification and implementation of a scalable analysis framework for ecmascript. In *FOOL 2012: 19th International Workshop on Foundations of Object-Oriented Languages*, page 96. Citeseer, 2012.

[82] Koushik Sen, Swaroop Kalasapur, Tasneem Brutch, and Simon Gibbs. Jalangi: A selective record-replay and dynamic analysis framework for javascript. In *Proceedings of the 2013 9th Joint Meeting on Foundations of Software Engineering*, pages 488–498, 2013.

[83] Esben Andreasen and Anders Møller. Determinacy in static analysis for jquery. In *Proceedings of the 2014 ACM International Conference on Object Oriented Programming Systems Languages & Applications*, pages 17–31, 2014.

[84] Changhee Park and Sukyoung Ryu. Scalable and precise static analysis of javascript applications via loop-sensitivity. In *29th European Conference on Object-Oriented Programming (ECOOP 2015)*. Schloss Dagstuhl-Leibniz-Zentrum fuer Informatik, 2015.

[85] Benno Stein, Benjamin Barslev Nielsen, Bor-Yuh Evan Chang, and Anders Møller. Static analysis with demand-driven value refinement. *Proceedings of the ACM on Programming Languages*, 3(OOPSLA):1–29, 2019.

[86] Joonyoung Park, Jihyeok Park, Dongjun Youn, and Sukyoung Ryu. Accelerating javascript static analysis via dynamic shortcuts. In *Proceedings of the 29th ACM Joint Meeting on European Software Engineering Conference and Symposium on the Foundations of Software Engineering*, pages 1129–1140, 2021.

[87] Cristian-Alexandru Staicu, Martin Toldam Torp, Max Schäfer, Anders Møller, and Michael Pradel. Extracting taint specifications for javascript libraries. In *Proceedings of the ACM/IEEE 42nd International Conference on Software Engineering*, pages 198–209, 2020.

[88] Cristian-Alexandru Staicu, Sazzadur Rahaman, Ágnes Kiss, and Michael Backes. Bilingual problems: Studying the security risks incurred by native extensions in scripting languages. *arXiv preprint arXiv:2111.11169*, 2021.

## Table 5: HODOR granularity of packages at system call level and thread level (RQ1).

| Attack Type | CVE | Package Name | Node.js with Musl Libc | | Hodor | | Node.js with Musl Libc | | Hodor | | # of CL | % of CL-1 | % of CL-2 |
|---|---|---|---|---|---|---|---|---|---|---|---|---|---|
| | | | # of CS | # of TS | # of CS | # of TS | # of MT | # of TP | # of MT | # of TP | | | |
| Arbitrary Command Injection | / | command-exists | 10 | 95 | 7 | 74 | 105 | 87 | 80 | 34 | 47 | 74.60% | 74.60% |
| Arbitrary Command Injection | CVE-2021-23363 | kill-by-port | 10 | 93 | 3 | 57 | 103 | 0 | 60 | 0 | 6 | 0.01% | 85.71% |
| Arbitrary Command Injection | CVE-2021-23360 | killport‡ | 11 | 99 | 11 | 72 | 110 | 0 | 83 | 0 | 237 | 18.53% | 95.83% |
| Arbitrary Command Injection | CVE-2021-23356 | kill-process-by-name | 11 | 99 | 11 | 72 | 110 | 0 | 83 | 0 | 6 | 75.00% | 75.00% |
| Arbitrary Command Injection | CVE-2018-13797 | macaddress | 10 | 93 | 5 | 58 | 103 | 0 | 63 | 0 | 52 | 36.36% | 36.36% |
| Arbitrary Command Injection | CVE-2022-25973 | mc-kill-port | 10 | 95 | 7 | 74 | 105 | 87 | 80 | 33 | 405 | 34.40% | 72.72% |
| Arbitrary Command Injection | CVE-2021-23377 | onion-oled-js | 10 | 93 | 3 | 57 | 103 | 0 | 60 | 0 | 145 | 10.11% | 82.35% |
| Arbitrary Command Injection | / | open | 10 | 93 | 3 | 57 | 103 | 0 | 60 | 0 | 12 | 47.82% | 47.82% |
| Arbitrary Command Injection | CVE-2018-3757 | pdf-image‡ | 11 | 99 | 11 | 75 | 110 | 87 | 85 | 34 | 194 | 25.19% | 86.15% |
| Arbitrary Command Injection | CVE-2018-3746 | pdfinfojs | 10 | 93 | 4 | 60 | 103 | 87 | 62 | 34 | 535 | 15.06% | 100.00% |
| Arbitrary Command Injection | CVE-2017-1000220 | pidusage‡ | 10 | 93 | 3 | 57 | 103 | 0 | 60 | 0 | 68 | 62.96% | 62.96% |
| Arbitrary Command Injection | CVE-2021-23379 | portkiller‡ | 10 | 93 | 3 | 57 | 103 | 0 | 60 | 0 | 12 | 88.23% | 88.23% |
| Arbitrary Command Injection | CVE-2021-23359 | port-killer | 10 | 95 | 7 | 72 | 105 | 0 | 79 | 0 | 8 | 100.00% | 100.00% |
| Arbitrary Command Injection | CVE-2021-23348 | portprocesses‡ | 10 | 93 | 3 | 57 | 103 | 0 | 60 | 0 | 19 | 88.57% | 88.57% |
| Arbitrary Command Injection | CVE-2018-16460 | ps | 10 | 93 | 3 | 57 | 103 | 0 | 60 | 0 | 28 | 75.00% | 75.00% |
| Arbitrary Command Injection | CVE-2021-23355 | ps-kill | 11 | 99 | 11 | 72 | 110 | 0 | 83 | 0 | 3 | 12.50% | 100.00% |
| Arbitrary Command Injection | CVE-2021-23374 | ps-visitor | 11 | 99 | 11 | 72 | 110 | 0 | 83 | 0 | 29 | 14.79% | 76.68% |
| Arbitrary Command Injection | CVE-2021-23380 | roar-pidusage‡ | 10 | 93 | 3 | 57 | 103 | 0 | 60 | 0 | 62 | 50.81% | 70.81% |
| Arbitrary Command Injection | / | samsung-remote | 11 | 99 | 11 | 75 | 110 | 87 | 85 | 34 | 27 | 56.52% | 56.52% |
| Arbitrary Command Injection | / | scp | 11 | 99 | 11 | 72 | 110 | 0 | 83 | 0 | 14 | 87.50% | 87.50% |
| Arbitrary Command Injection | CVE-2018-3772 | whereis‡ | 11 | 99 | 11 | 72 | 110 | 0 | 83 | 0 | 15 | 28.30% | 83.33% |
| Arbitrary Command Injection | CVE-2021-23399 | wincred | 11 | 99 | 11 | 72 | 110 | 0 | 83 | 0 | 10 | 76.92% | 76.92% |
| Argument Injection | CVE-2022-24437 | git-pull-or-clone | 10 | 95 | 9 | 75 | 105 | 87 | 82 | 36 | 177 | 31.72% | 92.85% |
| Command Injection | CVE-2020-7636 | adb-driver | 11 | 99 | 11 | 73 | 110 | 87 | 83 | 33 | 66 | 39.75% | 91.30% |
| Command Injection | / | alfred-workflow-nodejs | 10 | 93 | 4 | 60 | 103 | 87 | 62 | 34 | 729 | 2.18% | 86.95% |
| Command Injection | CVE-2018-16462 | apex-publish-static-files | 10 | 95 | 9 | 73 | 105 | 0 | 82 | 0 | 8 | 7.76% | 7.76% |
| Command Injection | CVE-2020-7633 | apiconnect-cli-plugins‡ | 10 | 93 | 3 | 57 | 103 | 0 | 60 | 0 | 19,230 | 8.83% | 11.43% |
| Command Injection | CVE-2021-3190 | async-git ‡ | 11 | 99 | 11 | 72 | 110 | 0 | 83 | 0 | 59 | 69.41% | 74.19% |
| Command Injection | CVE-2020-7730 | bestzip‡ | 11 | 99 | 11 | 82 | 110 | 87 | 92 | 35 | 4,204 | 12.60% | 89.04% |
| Command Injection | CVE-2019-10807 | blamer‡ | 10 | 93 | 3 | 57 | 103 | 0 | 60 | 0 | 1,465 | 13.88% | 57.57% |
| Command Injection | CVE-2020-7795 | cd-messenger | 10 | 95 | 9 | 72 | 105 | 0 | 81 | 0 | 374 | 36.77% | 83.33% |
| Command Injection | CVE-2020-7613 | clamscan‡ | 10 | 93 | 6 | 56 | 103 | 87 | 57 | 35 | 2,493 | 8.16% | 21.28% |
| Command Injection | / | cocos-utils | 10 | 93 | 4 | 64 | 103 | 87 | 66 | 34 | 95 | 9.17% | 9.17% |
| Command Injection | CVE-2020-15123 | codecov | 10 | 95 | 7 | 76 | 105 | 87 | 81 | 37 | 1,895 | 8.53% | 8.53% |
| Command Injection | CVE-2020-7635 | compass-compile | 10 | 93 | 3 | 57 | 103 | 0 | 60 | 0 | 202 | 13.96% | 95.65% |
| Command Injection | CVE-2020-7781TAB | connection-tester | 10 | 95 | 9 | 72 | 105 | 0 | 81 | 0 | 43 | 68.25% | 67.74% |
| Command Injection | CVE-2019-10789 | curling | 11 | 99 | 11 | 72 | 110 | 0 | 83 | 0 | 52 | 85.24% | 85.24% |
| Command Injection | CVE-2020-28425 | curljs | 11 | 99 | 11 | 72 | 110 | 0 | 83 | 0 | 75 | 66.37% | 78.12% |
| Command Injection | CVE-2020-28438 | deferred-exec | 11 | 99 | 11 | 72 | 110 | 0 | 83 | 0 | 1,339 | 9.93% | 87.75% |
| Command Injection | / | diskstats‡ | 10 | 93 | 3 | 57 | 103 | 0 | 60 | 0 | 55 | 94.82% | 94.82% |
| Command Injection | CVE-2020-7631 | diskusage-ng | 11 | 99 | 11 | 72 | 110 | 0 | 83 | 0 | 23 | 13.93% | 32.39% |
| Command Injection | CVE-2020-7606 | docker-compose-remote-api | 11 | 99 | 9 | 75 | 110 | 87 | 83 | 35 | 1,143 | 10.99% | 88.88% |
| Command Injection | CVE-2019-10801 | enpeem | 11 | 99 | 11 | 72 | 110 | 0 | 83 | 0 | 84 | 65.62% | 87.80% |
| Command Injection | CVE-2021-26275 | eslint-fixer‡ | 11 | 99 | 11 | 72 | 110 | 0 | 83 | 0 | 5,629 | 10.41% | 66.66% |
| Command Injection | CVE-2021-23376 | ffmpegdotjs ‡ | 11 | 99 | 11 | 73 | 110 | 0 | 84 | 0 | 43 | 41.74% | 36.95% |
| Command Injection | CVE-2021-23376 | ffmpeg-sdk ‡ | 11 | 99 | 11 | 72 | 110 | 0 | 83 | 0 | 18 | 75.00% | 75.00% |
| Command Injection | / | find-process‡ | 10 | 93 | 3 | 57 | 103 | 0 | 60 | 0 | 110 | 5.56% | 42.96% |
| Command Injection | / | freespace‡ | 11 | 101 | 11 | 80 | 112 | 0 | 91 | 0 | 27 | 44.26% | 59.09% |
| Command Injection | CVE-2020-28429 | geojson2kml | 10 | 93 | 3 | 57 | 103 | 0 | 60 | 0 | 6 | 100.00% | 100.00% |
| Command Injection | CVE-2020-7630 | git-add-remote | 11 | 101 | 11 | 80 | 112 | 0 | 91 | 0 | 13 | 65.00% | 65.00% |
| Command Injection | CVE-2020-28434 | gitblame | 11 | 99 | 11 | 72 | 110 | 0 | 83 | 0 | 9 | 18.00% | 75.00% |
| Command Injection | CVE-2018-3785 | git-dummy-commit | 11 | 101 | 11 | 87 | 112 | 0 | 98 | 0 | 490 | 7.41% | 88.88% |
| Command Injection | CVE-2019-10802 | giting | 11 | 99 | 11 | 73 | 110 | 87 | 83 | 34 | 197 | 20.02% | 64.56% |
| Command Injection | CVE-2022-1440 | git-interface | 10 | 93 | 9 | 60 | 103 | 0 | 69 | 0 | 94 | 68.11% | 68.11% |
| Command Injection | / | git-tags-remote‡ | 11 | 99 | 11 | 72 | 110 | 0 | 83 | 0 | 236 | 32.14% | 100.00% |
| Command Injection | CVE-2020-28436 | google-cloudstorage-commands‡ | 10 | 93 | 3 | 57 | 103 | 0 | 60 | 0 | 13 | 52.00% | 52.00% |
| Command Injection | CVE-2017-16042 | growl | 10 | 93 | 9 | 61 | 103 | 0 | 70 | 0 | 35 | 45.45% | 45.45% |
| Command Injection | CVE-2020-36650 | gry | 11 | 99 | 11 | 74 | 110 | 87 | 83 | 34 | 190 | 59.00% | 88.23% |
| Command Injection | CVE-2020-7601 | gulp-scss-lint‡ | 11 | 99 | 9 | 86 | 110 | 87 | 89 | 42 | 5,065 | 9.60% | 40.54% |
| Command Injection | CVE-2020-7607 | gulp-styledocco | 11 | 99 | 11 | 86 | 110 | 87 | 91 | 44 | 664 | 1.30% | 77.08% |
| Command Injection | CVE-2020-7605 | gulp-tape | 11 | 99 | 11 | 86 | 110 | 87 | 91 | 44 | 32 | 5.08% | 84.21% |
| Command Injection | CVE-2020-28437 | heroku-env | 11 | 99 | 11 | 75 | 110 | 87 | 85 | 34 | 25 | 33.33% | 42.00% |
| Command Injection | CVE-2019-10788 | im-metadata | 11 | 99 | 11 | 72 | 110 | 0 | 83 | 0 | 57 | 86.36% | 83.72% |
| Command Injection | CVE-2019-10787 | im-resize | 10 | 93 | 3 | 57 | 103 | 0 | 60 | 0 | 85 | 19.63% | 85.71% |
| Command Injection | CVE-2020-7629 | install-package‡ | 11 | 99 | 11 | 72 | 110 | 0 | 83 | 0 | 42 | 93.33% | 93.33% |
| Command Injection | CVE-2020-8178 | jison | 10 | 93 | 3 | 57 | 103 | 0 | 60 | 0 | 1,916 | 9.01% | 61.24% |
| Command Injection | CVE-2021-23381 | killing‡ | 11 | 99 | 11 | 72 | 110 | 0 | 83 | 0 | 38 | 9.76% | 90.24% |
| Command Injection | CVE-2019-15609 | kill-port-process | 11 | 101 | 11 | 81 | 112 | 87 | 91 | 33 | 259 | 20.45% | 56.75% |
| Command Injection | CVE-2018-16461 | libnmap | 11 | 99 | 11 | 74 | 110 | 87 | 83 | 34 | 3,069 | 6.30% | 80.93% |
| Command Injection | / | local-devices | 10 | 93 | 8 | 79 | 103 | 87 | 66 | 57 | 37 | 37.37% | 55.73% |
| Command Injection | CVE-2019-10783 | lsof | 11 | 99 | 11 | 72 | 110 | 0 | 83 | 0 | 54 | 91.30% | 91.30% |
| Command Injection | / | lycwed-spritesheetjs | 10 | 93 | 4 | 61 | 103 | 87 | 63 | 35 | 1,715 | 34.02% | 63.54% |
| Command Injection | CVE-2020-7786 | macfromip | 11 | 99 | 11 | 75 | 110 | 87 | 85 | 34 | 39 | 45.88% | 45.88% |
| Command Injection | CVE-2020-28434 | monorepo-build | 10 | 95 | 7 | 72 | 105 | 0 | 79 | 0 | 7,810 | 7.58% | 85.18% |
| Command Injection | CVE-2019-10786 | network-manager | 10 | 95 | 9 | 72 | 105 | 0 | 81 | 0 | 93 | 89.42% | 93.00% |
| Command Injection | CVE-2019-15597 | node-df | 11 | 99 | 11 | 72 | 110 | 0 | 83 | 0 | 84 | 1.71% | 91.30% |
| Command Injection | CVE-2020-7627 | node-key-sender‡ | 11 | 99 | 11 | 72 | 110 | 0 | 83 | 0 | 93 | 77.50% | 77.50% |
| Command Injection | CVE-2020-28433 | node-latex-pdf | 11 | 99 | 11 | 72 | 110 | 0 | 83 | 0 | 9 | 60.00% | 60.00% |
| Command Injection | CVE-2020-7632 | node-mpv | 10 | 95 | 7 | 72 | 105 | 0 | 79 | 0 | 36 | 14.63% | 14.63% |
| Command Injection | CVE-2020-7602 | node-prompt-here | 10 | 95 | 7 | 72 | 105 | 0 | 79 | 0 | 8 | 6.66% | 57.14% |
| Command Injection | CVE-2020-7785 | node-ps | 11 | 99 | 11 | 72 | 110 | 0 | 83 | 0 | 37 | 75.51% | 75.51% |
| Command Injection | / | node-unrar | 11 | 99 | 11 | 72 | 110 | 0 | 83 | 0 | 17 | 8.21% | 100.00% |
| Command Injection | CVE-2022-0841 | npm-lockfile‡ | 10 | 93 | 3 | 57 | 103 | 0 | 60 | 0 | 1,963 | 7.01% | 70.96% |
| Command Injection | CVE-2021-23375 | psnode | 11 | 99 | 11 | 72 | 110 | 0 | 83 | 0 | 63 | 6.12% | 35.13% |
| Command Injection | CVE-2020-7604 | pulverizr | 11 | 99 | 11 | 77 | 110 | 87 | 87 | 33 | 614 | 6.56% | 87.02% |
| Command Injection | CVE-2021-24033 | react-dev-utils | 10 | 95 | 9 | 72 | 105 | 0 | 81 | 0 | 6,425 | 11.77% | 22.32% |
| Command Injection | CVE-2019-10796 | rpi | 11 | 99 | 11 | 72 | 110 | 0 | 83 | 0 | 28 | 65.11% | 65.00% |
| Command Injection | CVE-2019-10804 | serial-number | 11 | 99 | 11 | 73 | 110 | 87 | 83 | 33 | 45 | 57.69% | 57.69% |
| Command Injection | / | strider-git | 11 | 99 | 11 | 77 | 110 | 87 | 87 | 36 | 464 | 24.82% | 42.40% |
| Command Injection | CVE-2020-7621 | strong-nginx-controller | 10 | 93 | 3 | 57 | 103 | 0 | 60 | 0 | 17,994 | 6.37% | 49.70% |
| Command Injection | CVE-2020-28432 | theme-core‡ | 10 | 93 | 4 | 58 | 103 | 87 | 60 | 33 | 1,721 | 15.46% | 81.35% |
| Command Injection | CVE-2020-7784 | ts-process-promises | 11 | 99 | 11 | 73 | 110 | 87 | 83 | 33 | 2,376 | 16.29% | 84.76%% |
| Command Injection | CVE-2020-7628 | umount | 11 | 99 | 11 | 72 | 110 | 0 | 83 | 0 | 965 | 8.20% | 71.73% |
| Command Injection | / | vboxmanage.js‡ | 11 | 99 | 11 | 75 | 110 | 87 | 85 | 34 | 55 | 25.82% | 25.82% |
| Command Injection | CVE-2020-28431 | wc-cmd‡ | 11 | 99 | 11 | 72 | 110 | 0 | 83 | 0 | 31 | 0.44% | 82.50% |
| Command Injection | CVE-2020-15362 | wifiscanner | 11 | 99 | 11 | 73 | 110 | 87 | 83 | 33 | 215 | 9.10% | 83.33% |

| | | | | | | | | | | | | |
|---|---|---|---|---|---|---|---|---|---|---|---|---|
| **Command Injection** | **CVE-2020-28447** | **xopen‡** | 11 | 99 | 11 | 72 | 110 | 0 | 83 | 0 | 9 | 90.00% | 90.00% |
| **Command Injection** | **/** | **xps** | 11 | 99 | 11 | 72 | 110 | 0 | 83 | 0 | 51 | 4.26% | 54.54% |
| **Remote Code Execution** | **CVE-2020-36378** | **aaptjs** | 11 | 101 | 11 | 87 | 112 | 0 | 98 | 0 | 457 | 13.00% | 97.05% |
| **Remote Code Execution** | **/** | **arpping‡** | 11 | 99 | 11 | 72 | 110 | 0 | 83 | 0 | 98 | 27.68% | 78.40% |
| **Remote Code Execution** | **CVE-2020-11079** | **dns-sync‡** | 11 | 101 | 11 | 83 | 112 | 87 | 93 | 35 | 344 | 27.80% | 97.22% |
| **Remote Code Execution** | **CVE-2021-23632** | **git** | 10 | 93 | 3 | 57 | 103 | 0 | 60 | 0 | 648 | 15.72% | 15.89% |
| **Remote Code Execution** | **/** | **git-lib‡** | 10 | 93 | 3 | 57 | 103 | 0 | 60 | 0 | 999 | 11.35% | 59.02% |
| **Remote Code Execution** | **/** | **git-parse** | 10 | 93 | 4 | 60 | 103 | 87 | 62 | 34 | 704 | 2.96% | 39.51% |
| **Remote Code Execution** | **/** | **gity** | 11 | 99 | 11 | 72 | 110 | 0 | 83 | 0 | 33 | 13.25% | 55.93% |
| **Remote Code Execution** | **/** | **imagickal‡** | 11 | 99 | 11 | 75 | 110 | 87 | 85 | 35 | 1,183 | 13.13% | 83.16% |
| **Remote Code Execution** | **/** | **meta-git‡** | 10 | 93 | 3 | 57 | 103 | 0 | 60 | 0 | 355 | 8.32% | 89.02% |
| **Remote Code Execution** | **/** | **node-os-utils‡** | 10 | 95 | 7 | 72 | 105 | 0 | 79 | 0 | 172 | 47.77% | 47.77% |
| **Remote Code Execution** | **CVE-2020-7620** | **pomelo-monitor** | 10 | 93 | 3 | 58 | 103 | 0 | 61 | 0 | 90 | 86.53% | 86.53% |
| **Remote Shell Command Injection** | **CVE-2015-7982** | **gm‡** | 10 | 93 | 3 | 57 | 103 | 0 | 60 | 0 | 675 | 49.52% | 58.50% |
| **Arbitrary Code Execution** | **CVE-2020-7729** | **grunt** | 11 | 98 | 5 | 67 | 109 | 87 | 70 | 33 | 6,747 | 7.69% | 47.13% |
| **Arbitrary Code Execution** | **/** | **is-my-json-valid** | 1 | 28 | 1 | 28 | 29 | 0 | 29 | 0 | 3,560 | 9.92% | 93.84% |
| **Arbitrary Code Execution** | **CVE-2020-7777** | **jsen** | 11 | 99 | 11 | 72 | 110 | 0 | 83 | 0 | 766 | 34.56% | 95.15% |
| **Arbitrary Code Execution** | **CVE-2020-7673** | **node-extend** | 1 | 28 | 1 | 28 | 29 | 0 | 29 | 0 | 24 | 82.75% | 82.75% |
| **Arbitrary Code Execution** | **CVE-2017-16082** | **pg** | 1 | 28 | 1 | 28 | 29 | 0 | 29 | 0 | 260 | 12.26% | 27.54% |
| **Arbitrary Code Execution** | **CVE-2020-7640** | **pixl-class** | 1 | 28 | 1 | 28 | 29 | 0 | 29 | 0 | 38 | 79.16% | 79.16% |
| **Arbitrary Code Execution** | **CVE-2022-0748** | **post-loader** | 6 | 83 | 1 | 40 | 89 | 0 | 41 | 0 | 900 | 8.64% | 69.23% |
| **Arbitrary Code Execution** | **/** | **serialize-to-js** | 1 | 28 | 1 | 28 | 29 | 0 | 29 | 0 | 120 | 5.10% | 86.33% |
| **Arbitrary Code Execution** | **CVE-2021-23389** | **total.js** | 6 | 83 | 2 | 46 | 89 | 87 | 44 | 35 | 2,072 | 6.55% | 6.55% |
| **Arbitrary Code Execution** | **CVE-2021-23390** | **total4** | 10 | 94 | 5 | 72 | 104 | 87 | 75 | 38 | 1,874 | 6.79% | 6.79% |
| **Arbitrary Code Execution** | **CVE-2017-1001004** | **typed-function** | 1 | 28 | 1 | 28 | 29 | 0 | 29 | 0 | 397 | 74.62% | 74.62% |
| **Arbitrary Code Injection** | **/** | **kmc** | 6 | 83 | 1 | 45 | 89 | 0 | 46 | 0 | 2,267 | 6.02% | 55.81% |
| **Arbitrary Code Injection** | **/** | **marsdb** | 1 | 28 | 1 | 28 | 29 | 0 | 29 | 0 | 1,825 | 18.20% | 54.06% |
| **Arbitrary Code Injection** | **/** | **mixin-pro** | 1 | 28 | 1 | 28 | 29 | 0 | 29 | 0 | 78 | 82.10% | 82.10% |
| **Arbitrary Code Injection** | **/** | **m-log** | 1 | 28 | 1 | 28 | 29 | 0 | 29 | 0 | 289 | 0.88% | 100.00% |
| **Arbitrary Code Injection** | **/** | **mobile-icon-resizer** | 10 | 93 | 4 | 59 | 103 | 87 | 61 | 35 | 54 | 33.75% | 65.51% |
| **Arbitrary Code Injection** | **/** | **mock2easy** | 10 | 93 | 5 | 70 | 103 | 87 | 73 | 35 | / | / | / |
| **Arbitrary Code Injection** | **/** | **modjs** | 10 | 93 | 4 | 62 | 103 | 87 | 64 | 34 | 167 | 2.11% | 6.77% |
| **Arbitrary Code Injection** | **/** | **modulify** | 6 | 83 | 1 | 40 | 89 | 0 | 41 | 0 | 1,133 | 1.97% | 80.70% |
| **Arbitrary Code Injection** | **/** | **mol-proto** | 1 | 28 | 1 | 28 | 29 | 0 | 29 | 0 | 403 | 90.76% | 90.76% |
| **Arbitrary Code Injection** | **/** | **mongoosemask** | 1 | 28 | 1 | 28 | 29 | 0 | 29 | 0 | 50 | 19.92% | 67.56% |
| **Arbitrary Code Injection** | **/** | **protojs** | 1 | 28 | 1 | 28 | 29 | 0 | 29 | 0 | 132 | 10.09% | 90.76% |
| **Arbitrary Code Injection** | **CVE-2020-7660** | **serialize-javascript** | 1 | 28 | 1 | 28 | 29 | 0 | 29 | 0 | 71 | 5.10% | 93.42% |
| **Arbitrary File Overwrite** | **CVE-2021-32803** | **tar** | 6 | 83 | 3 | 56 | 89 | 87 | 57 | 48 | 7,830 | 11.69% | 83.14% |
| **Arbitrary File Write** | **CVE-2018-1002204** | **adm-zip‡** | 6 | 83 | 1 | 41 | 89 | 0 | 42 | 0 | 516 | 26.81% | 26.84% |
| **Code Execution** | **CVE-2017-5941** | **node-serialize** | 1 | 28 | 1 | 28 | 29 | 0 | 29 | 0 | 52 | 92.85% | 92.85% |
| **Code Injection** | **CVE-2022-25760** | **accesslog** | 6 | 83 | 1 | 43 | 89 | 0 | 44 | 0 | 95 | 41.48% | 83.33% |
| **Code Injection** | **CVE-2020-7674** | **access-policy** | 1 | 28 | 1 | 28 | 29 | 0 | 29 | 0 | 81 | 0.26% | 96.42% |
| **Code Injection** | **CVE-2021-21277** | **angular-expressions** | 1 | 28 | 1 | 28 | 29 | 0 | 29 | 0 | 713 | 48.01% | 48.01% |
| **Code Injection** | **CVE-2020-7675** | **cd-messenger** | 1 | 28 | 1 | 28 | 29 | 0 | 29 | 0 | 374 | 36.77% | 90.69% |
| **Code Injection** | **CVE-2018-3784** | **cryop** | 1 | 28 | 1 | 28 | 29 | 0 | 29 | 0 | 90 | 84.11% | 84.11% |
| **Code Injection** | **CVE-2019-15657** | **eslint-utils** | 6 | 83 | 2 | 50 | 89 | 87 | 48 | 35 | 436 | 8.92% | 80.89% |
| **Code Injection** | **CVE-2021-23639** | **front-matter** | 6 | 83 | 2 | 42 | 89 | 87 | 36 | 33 | 1,096 | 11.35% | 80.95% |
| **Code Injection** | **CVE-2020-6836** | **hot-formula-parser** | 1 | 28 | 1 | 28 | 29 | 0 | 29 | 0 | 1,650 | 6.25% | 89.94% |
| **Code Injection** | **/** | **js-yaml** | 1 | 28 | 1 | 28 | 29 | 0 | 29 | 0 | 1,753 | 19.62% | 70.66% |
| **Code Injection** | **CVE-2022-21122** | **metacalc** | 7 | 90 | 2 | 52 | 97 | 0 | 54 | 0 | 59 | 54.12% | 86.95% |
| **Code Injection** | **CVE-2019-5413** | **morgan** | 6 | 83 | 2 | 48 | 89 | 87 | 46 | 35 | 2,135 | 11.69% | 81.48% |
| **Code Injection** | **CVE-2022-25921** | **morgan-json** | 6 | 83 | 1 | 32 | 89 | 0 | 33 | 0 | 261 | 12.59% | 94.11% |
| **Code Injection** | **CVE-2020-7672** | **mosc.js** | 1 | 28 | 1 | 28 | 29 | 0 | 29 | 0 | 47 | 0.93% | 88.67% |
| **Code Injection** | **CVE-2020-7609** | **node-rules.js** | 1 | 28 | 1 | 28 | 29 | 0 | 29 | 0 | 939 | 40.57% | 91.58% |
| **Code Injection** | **CVE-2016-10548** | **reduce-css-calc** | 1 | 28 | 1 | 28 | 29 | 0 | 29 | 0 | 110 | 87.27% | 87.27% |
| **Code Injection** | **CVE-2020-7677** | **thenify** | 1 | 28 | 1 | 28 | 29 | 0 | 29 | 0 | 50 | 56.81% | 82.14% |
| **Prototype Pollution** | **CVE-2020-7743** | **mathjs** | 1 | 28 | 1 | 28 | 29 | 0 | 29 | 0 | 2,650 | 9.83% | 8.52% |
| **Prototype Pollution** | **CVE-2021-23594** | **realms-shim** | 6 | 83 | 1 | 38 | 89 | 0 | 39 | 0 | 1,118 | 3.59% | 78.18% |
| **Remote Code Execution** | **/** | **djv** | 1 | 28 | 1 | 28 | 29 | 0 | 29 | 0 | 300 | 33.44% | 55.35% |
| **Remote Code Execution** | **/** | **mongodb-query-parser** | 7 | 90 | 3 | 60 | 97 | 87 | 59 | 35 | 2,688 | 5.32% | 28.49% |
| **Remote Code Execution** | **CVE-2019-10758** | **mongo-express** | 6 | 83 | 2 | 43 | 89 | 87 | 42 | 33 | 9,225 | 11.46% | 23.58% |
| **Remote Code Execution** | **CVE-2020-24391** | **mongo-parse** | 1 | 28 | 1 | 28 | 29 | 0 | 29 | 0 | 342 | 83.82% | 83.82% |
| **Sandbox Breakout** | **/** | **notevil** | 1 | 28 | 1 | 28 | 29 | 0 | 29 | 0 | 2,828 | 9.68% | 89.58% |
| **Sandbox Breakout** | **CVE-2019-10769** | **safer-eval** | 7 | 90 | 2 | 52 | 97 | 0 | 54 | 0 | 119 | 35.31% | 80.76% |
| **Sandbox Breakout** | **/** | **sandbox** | 10 | 93 | 4 | 58 | 103 | 87 | 60 | 33 | 42 | 48.88% | 48.88% |
| **Sandbox Breakout** | **/** | **static-eval** | 11 | 99 | 9 | 74 | 110 | 87 | 82 | 34 | 2,891 | 7.69% | 75.51% |
| **Sandbox Breakout** | **/** | **value-censorship** | 7 | 90 | 2 | 52 | 97 | 0 | 54 | 0 | 3,089 | 8.43% | 95.00% |
| **Sandbox Bypass** | **CVE-2019-10761** | **vm2** | 7 | 90 | 2 | 52 | 97 | 0 | 54 | 0 | 50 | 0.09% | 53.33% |
| **Sandbox Escape** | **CVE-2020-7710** | **safe-eval** | 6 | 83 | 1 | 38 | 89 | 0 | 39 | 0 | 11 | 95.15% | 95.15% |
| **Sandbox Escape** | **CVE-2020-7710** | **zhaoyao91-eval-in-vm** | 6 | 83 | 1 | 38 | 89 | 0 | 39 | 0 | 7 | 100.00% | 100.00% |
| **Template Injection** | **CVE-2022-29078** | **ejs** | 6 | 83 | 1 | 40 | 89 | 0 | 41 | 0 | 32 | 5.22% | 87.33% |
| **Template Injection** | **CVE-2021-23358** | **underscore** | 1 | 28 | 1 | 28 | 29 | 0 | 29 | 0 | 239 | 48.38% | 48.38% |
| **Arbitrary Command Execution** | | | **1,161** | **10,636** | **910** | **7,617** | **11,797** | **2,958** | **8,444** | **1,212** | **936** | 35.61% | 69.14% |
| **Arbitrary Code Execution** | | | **243** | **3,352** | **110** | **2,294** | **3,595** | **1,218** | **2,362** | **496** | **1,178** | 30.83% | 71.03% |

**CS**: Critical system calls invocation;
**TS**: Trivial system calls invocation;
**MT**: Main Thread system calls invocation;
**TP**: Thread Pool system calls invocation;
‡: Packages that invoke system calls that are triggered by the execution of builtin methods;
**CL**: Covered line num of the package;
**CL-1**: Covered line of the package;
**CL-2**: Covered line of the package module;

## Table 6: Exploit execution for packages with arbitrary command execution attacks.

| Package Name | Initial Attack | HODOR | | | | | | | |
|---|---|---|---|---|---|---|---|---|---|
| | | Initial Attack | Cmd | | Permission | | Net | | |
| | | | Exec | Fork | Setgid | Setuid | Connect | Listen | Bind |
| command-exists | Write command-exists | ✔ | ✘ | ✘ | ✔ | ✔ | ✘ | ✔ | ✔ |
| kill-by-port | Write success | ✔ | ✘ | ✘ | ✔ | ✔ | ✔ | ✔ | ✔ |
| killport | Write success | ✔ | ✘ | ✘ | ✔ | ✔ | ✔ | ✔ | ✔ |
| kill-process-by-name | Write success | ✔ | ✘ | ✘ | ✔ | ✔ | ✔ | ✔ | ✔ |
| macaddress | Write /tmp/poof | ✔ | ✘ | ✘ | ✔ | ✔ | ✔ | ✔ | ✔ |
| mc-kill-port | Write newFile.txt | ✔ | ✘ | ✘ | ✔ | ✔ | ✘ | ✔ | ✔ |
| onion-oled-js | Write success | ✔ | ✘ | ✘ | ✔ | ✔ | ✔ | ✔ | ✔ |
| open | Write /tmp/tada | ✔ | ✘ | ✘ | ✔ | ✔ | ✔ | ✔ | ✔ |
| pdf-image | Write /tmp/hacked | ✔ | ✘ | ✘ | ✔ | ✔ | ✔ | ✔ | ✔ |
| pdfinfojs | Write a | ✔ | ✘ | ✘ | ✔ | ✔ | ✔ | ✔ | ✔ |
| pidusage | Execute /usr/local/bin/python | ✔ | ✘ | ✘ | ✔ | ✔ | ✔ | ✔ | ✔ |
| portkiller | Write success | ✔ | ✘ | ✘ | ✔ | ✔ | ✔ | ✔ | ✔ |
| port-killer | Write success | ✔ | ✘ | ✘ | ✔ | ✔ | ✘ | ✔ | ✔ |
| portprocesses | Write success | ✔ | ✘ | ✘ | ✔ | ✔ | ✔ | ✔ | ✔ |
| ps | Write success.txt | ✔ | ✘ | ✘ | ✔ | ✔ | ✔ | ✔ | ✔ |
| ps-kill | Write success | ✔ | ✘ | ✘ | ✔ | ✔ | ✔ | ✔ | ✔ |
| ps-visitor | Write success | ✔ | ✘ | ✘ | ✔ | ✔ | ✔ | ✔ | ✔ |
| roar-pidusage | Write success | ✔ | ✘ | ✘ | ✔ | ✔ | ✔ | ✔ | ✔ |
| samsung-remote | Write /tmp/malicious; | ✔ | ✘ | ✘ | ✔ | ✘ | ✔ | ✔ | ✔ |
| scp | nc localhost 4444; | ✔ | ✘ | ✘ | ✔ | ✔ | ✔ | ✔ | ✔ |
| whereis | Write /tmp/tada | ✔ | ✘ | ✘ | ✔ | ✔ | ✔ | ✔ | ✔ |
| wincred | Write success | ✔ | ✘ | ✘ | ✔ | ✔ | ✔ | ✔ | ✔ |
| git-pull-or-clone | Write /tmp/pwn3 | ✔ | ✘ | ✘ | ✔ | ✔ | ✘ | ✔ | ✔ |
| adb-driver | Write a | ✔ | ✘ | ✘ | ✔ | ✔ | ✔ | ✔ | ✔ |
| alfred-workflow-nodejs | Write hacked | ✔ | ✘ | ✘ | ✔ | ✔ | ✔ | ✔ | ✔ |
| apex-publish-static-files | Write apex-publish-static-files | ✔ | ✘ | ✘ | ✔ | ✔ | ✘ | ✔ | ✔ |
| apiconnect-cli-plugins | Write Song | ✔ | ✘ | ✘ | ✔ | ✔ | ✔ | ✔ | ✔ |
| async-git | Write HACKED # | ✔ | ✘ | ✘ | ✘ | ✘ | ✘ | ✘ | ✘ |
| bestzip | Write bestzip | ✔ | ✘ | ✘ | ✔ | ✔ | ✔ | ✔ | ✔ |
| blamer | Write vulnerable | ✔ | ✘ | ✘ | ✔ | ✔ | ✔ | ✔ | ✔ |
| cd-messenger | Write JHU | ✔ | ✘ | ✘ | ✔ | ✔ | ✘ | ✔ | ✔ |
| clamscan | Write create.txt | ✔ | ✘ | ✘ | ✔ | ✔ | ✘ | ✔ | ✔ |
| cocos-utils | Write hacked | ✔ | ✘ | ✘ | ✔ | ✔ | ✔ | ✔ | ✔ |
| codecov | Write codecov | ✔ | ✘ | ✘ | ✔ | ✔ | ✘ | ✔ | ✔ |
| compass-compile | Write JHU | ✔ | ✘ | ✘ | ✔ | ✔ | ✔ | ✔ | ✔ |
| connection-tester | Write 111 | ✔ | ✘ | ✘ | ✔ | ✔ | ✘ | ✔ | ✔ |
| curling | Write JHU | ✔ | ✘ | ✘ | ✔ | ✔ | ✔ | ✔ | ✔ |
| curljs | Write JHU | ✔ | ✘ | ✘ | ✔ | ✔ | ✔ | ✔ | ✔ |
| deferred-exec | Write JHU | ✔ | ✘ | ✘ | ✔ | ✔ | ✔ | ✔ | ✔ |
| diskstats | Write HACKED | ✔ | ✘ | ✘ | ✔ | ✔ | ✔ | ✔ | ✔ |
| diskusage-ng | Write Song | ✔ | ✘ | ✘ | ✔ | ✔ | ✔ | ✔ | ✔ |
| docker-compose-remote-api | Write vulnerable.txt | ✔ | ✘ | ✘ | ✔ | ✔ | ✘ | ✘ | ✘ |
| enpeem | Write create.txt | ✔ | ✘ | ✘ | ✔ | ✔ | ✔ | ✔ | ✔ |
| eslint-fixer | Write eslint-fixer | ✔ | ✘ | ✘ | ✘ | ✘ | ✔ | ✔ | ✔ |
| ffmpegdotjs | Write success | ✔ | ✘ | ✘ | ✔ | ✔ | ✔ | ✔ | ✔ |
| ffmpeg-sdk | Write success | ✔ | ✘ | ✘ | ✔ | ✔ | ✔ | ✔ | ✔ |
| find-process | Write /tmp/semicolon_file | ✔ | ✘ | ✘ | ✔ | ✔ | ✔ | ✔ | ✔ |
| freespace | Write /tmp/semicolon_file | ✔ | ✘ | ✘ | ✔ | ✔ | ✘ | ✔ | ✔ |
| geojson2kml | Write JHU | ✔ | ✘ | ✘ | ✔ | ✔ | ✔ | ✔ | ✔ |
| git-add-remote | Write Song | ✔ | ✘ | ✘ | ✔ | ✔ | ✘ | ✔ | ✔ |
| gitblame | Write JHU | ✔ | ✘ | ✘ | ✔ | ✔ | ✔ | ✔ | ✔ |
| git-dummy-commit | Write git-dummy-commit | ✔ | ✘ | ✘ | ✔ | ✔ | ✘ | ✔ | ✔ |
| giting | Write create.txt | ✔ | ✘ | ✘ | ✔ | ✔ | ✔ | ✔ | ✔ |
| git-interface | Write /tmp/pwned | ✔ | ✘ | ✘ | ✔ | ✔ | ✔ | ✔ | ✔ |
| git-tags-remote | Write /tmp/command-injection.test | ✔ | ✘ | ✘ | ✔ | ✔ | ✔ | ✔ | ✔ |
| google-cloudstorage-commands | Write JHU | ✔ | ✘ | ✘ | ✔ | ✔ | ✔ | ✔ | ✔ |
| growl | Write aaaa | ✔ | ✘ | ✘ | ✔ | ✔ | ✔ | ✔ | ✔ |
| gry | Write HACKED | ✔ | ✘ | ✘ | ✔ | ✔ | ✔ | ✔ | ✔ |
| gulp-scss-lint | Write create.txt | ✔ | ✘ | ✘ | ✔ | ✔ | ✘ | ✘ | ✘ |
| gulp-styledocco | Write Vulnerable | ✔ | ✘ | ✘ | ✔ | ✔ | ✘ | ✘ | ✘ |
| gulp-tape | Write JHU.txt | ✔ | ✘ | ✘ | ✘ | ✘ | ✘ | ✘ | ✘ |
| heroku-env | Write JHU | ✔ | ✘ | ✘ | ✘ | ✘ | ✔ | ✔ | ✔ |
| im-metadata | Write im-metadata | ✔ | ✘ | ✘ | ✘ | ✘ | ✘ | ✘ | ✘ |
| im-resize | Write create.txt | ✔ | ✘ | ✘ | ✔ | ✔ | ✔ | ✔ | ✔ |
| install-package | Write Song | ✔ | ✘ | ✘ | ✔ | ✔ | ✔ | ✔ | ✔ |
| jison | Write pwned | ✔ | ✘ | ✘ | ✔ | ✔ | ✔ | ✔ | ✔ |
| killing | Write success | ✔ | ✘ | ✘ | ✔ | ✔ | ✔ | ✔ | ✔ |
| kill-port-process | Write kill-port-process | ✔ | ✘ | ✘ | ✔ | ✔ | ✘ | ✔ | ✔ |
| libnmap | Write success.txt | ✔ | ✘ | ✘ | ✔ | ✔ | ✔ | ✔ | ✔ |
| local-devices | Makek directory attacker | ✔ | ✘ | ✘ | ✔ | ✔ | ✔ | ✔ | ✔ |
| lsof | Write create.txt | ✔ | ✘ | ✘ | ✘ | ✘ | ✔ | ✔ | ✔ |
| lycwed-spritesheetjs | Write 111233 # | ✔ | ✘ | ✘ | ✔ | ✔ | ✔ | ✔ | ✔ |
| macfromip | Write JHU2 | ✔ | ✘ | ✘ | ✔ | ✘ | ✔ | ✔ | ✔ |
| monorepo-build | Write JHU | ✔ | ✘ | ✘ | ✔ | ✔ | ✘ | ✔ | ✔ |
| network-manager | Write create.txt | ✔ | ✘ | ✘ | ✔ | ✔ | ✘ | ✔ | ✔ |
| node-df | Write HACKED | ✔ | ✘ | ✘ | ✔ | ✔ | ✔ | ✔ | ✔ |
| node-key-sender | Write Song | ✔ | ✘ | ✘ | ✔ | ✔ | ✔ | ✔ | ✔ |
| node-latex-pdf | Write JHU | ✔ | ✘ | ✘ | ✘ | ✘ | ✘ | ✘ | ✘ |
| node-mpv | Write JHU | ✔ | ✘ | ✘ | ✔ | ✔ | ✘ | ✔ | ✔ |
| node-prompt-here | Write create.txt | ✔ | ✘ | ✘ | ✔ | ✔ | ✘ | ✔ | ✔ |
| node-ps | Write JHU | ✔ | ✘ | ✘ | ✔ | ✔ | ✔ | ✔ | ✔ |
| node-unrar | Write node-unrar | ✔ | ✘ | ✘ | ✔ | ✔ | ✔ | ✔ | ✔ |
| npm-lockfile | Write rce | ✔ | ✘ | ✘ | ✔ | ✔ | ✔ | ✔ | ✔ |
| psnode | Write success | ✔ | ✘ | ✘ | ✔ | ✔ | ✔ | ✔ | ✔ |
| pulverizr | Write Song | ✔ | ✘ | ✘ | ✔ | ✔ | ✔ | ✔ | ✔ |
| react-dev-utils | Write react-dev-utils | ✔ | ✘ | ✘ | ✔ | ✔ | ✘ | ✔ | ✔ |
| rpi | Write vulnerable.txt | ✔ | ✘ | ✘ | ✔ | ✔ | ✔ | ✔ | ✔ |
| serial-number | Write create.txt | ✔ | ✘ | ✘ | ✘ | ✘ | ✔ | ✔ | ✔ |
| strider-git | Write HACKED; | ✔ | ✘ | ✘ | ✘ | ✘ | ✘ | ✘ | ✘ |
| strong-nginx-controller | Write Song | ✔ | ✘ | ✘ | ✔ | ✔ | ✔ | ✔ | ✔ |
| theme-core | Write JHU | ✔ | ✘ | ✘ | ✔ | ✔ | ✔ | ✔ | ✔ |
| ts-process-promises | Write JHU | ✔ | ✘ | ✘ | ✔ | ✔ | ✔ | ✔ | ✔ |
| umount | Write Song | ✔ | ✘ | ✘ | ✘ | ✘ | ✔ | ✔ | ✔ |
| vboxmanage.js | Write HACKED | ✔ | ✘ | ✘ | ✔ | ✔ | ✔ | ✔ | ✔ |
| wc-cmd | Write JHU | ✔ | ✘ | ✘ | ✔ | ✔ | ✔ | ✔ | ✔ |
| wifiscanner | Write /tmp/exploit.txt | ✔ | ✘ | ✘ | ✔ | ✔ | ✔ | ✔ | ✔ |
| xopen | Write JHU | ✔ | ✘ | ✘ | ✔ | ✔ | ✔ | ✔ | ✔ |
| xps | Write HACKED | ✔ | ✘ | ✘ | ✔ | ✔ | ✔ | ✔ | ✔ |
| aaptjs | Write aaptjs | ✔ | ✘ | ✘ | ✘ | ✘ | ✘ | ✘ | ✘ |
| arpping | Write HACKED | ✔ | ✘ | ✘ | ✔ | ✔ | ✔ | ✔ | ✔ |
| dns-sync | Write pwned | ✔ | ✘ | ✘ | ✔ | ✔ | ✘ | ✔ | ✔ |
| git | date | ✘ | ✘ | ✘ | ✔ | ✔ | ✘ | ✔ | ✔ |
| git-lib | Write HACKED; | ✔ | ✘ | ✘ | ✔ | ✔ | ✔ | ✔ | ✔ |
| git-parse | Write HACKED | ✔ | ✘ | ✘ | ✔ | ✔ | ✔ | ✔ | ✔ |
| gity | Write HACKED | ✔ | ✘ | ✘ | ✔ | ✔ | ✔ | ✔ | ✔ |
| imagickal | Write HACKED | ✔ | ✘ | ✘ | ✔ | ✔ | ✔ | ✔ | ✔ |
| meta-git | Write HACKED | ✔ | ✘ | ✘ | ✔ | ✔ | ✔ | ✔ | ✔ |
| node-os-utils | Write DUMMY_FILE | ✔ | ✘ | ✘ | ✔ | ✔ | ✘ | ✔ | ✔ |
| pomelo-monitor | Write Song | ✔ | ✘ | ✘ | ✔ | ✔ | ✔ | ✔ | ✔ |
| gm | Write gm | ✔ | ✘ | ✘ | ✔ | ✔ | ✔ | ✔ | ✔ |

✘: Exploits are executed; ✔: Exploits are blocked;

## Table 7: Exploit execution for packages with arbitrary code execution attacks.

| Package Name | Initial Attack | HODOR | | | | | | | | MIR | | | | | | | |
|---|---|---|---|---|---|---|---|---|---|---|---|---|---|---|---|---|---|
| | | Initial Attack | Cmd | | Permission | | Net | | | Initial Attack | Cmd | | Permission | | Net | | |
| | | | Exec | Fork | Setgid | Setuid | Connect | Listen | Bind | | Exec | Fork | Setgid | Setuid | Connect | Listen | Bind |
| accesslog | Print xSS | ✘ | ✔ | ✔ | ✔ | ✔ | ✔ | ✔ | ✔ | / | | | | | | | |
| access-policy | Print 123 | ✘ | ✔ | ✔ | ✔ | ✔ | ✔ | ✔ | ✔ | ✔ | ✔ | ✔ | ✔ | ✔ | ✔ | ✔ | ✔ |
| adm-zip | Path traversal | ✔ | ✔ | ✔ | ✔ | ✔ | ✔ | ✔ | ✔ | / | | | | | | | |
| angular-expressions | Write file angular-expressions-success | ✔ | ✔ | ✔ | ✔ | ✔ | ✔ | ✔ | ✔ | / | | | | | | | |
| cd-messenger | Print JHU | ✘ | ✔ | ✔ | ✔ | ✔ | ✔ | ✔ | ✔ | ✔ | ✔ | ✔ | ✔ | ✔ | ✔ | ✔ | ✔ |
| cryop | Print defconrussia | ✘ | ✔ | ✔ | ✔ | ✔ | ✔ | ✔ | ✔ | ✔ | ✔ | ✔ | ✔ | ✔ | ✔ | ✔ | ✔ |
| djv | touch HACKED | ✔ | ✔ | ✔ | ✔ | ✔ | ✔ | ✔ | ✔ | / | | | | | | | |
| ejs | Write file ejs-success | ✔ | ✔ | ✔ | ✔ | ✔ | ✔ | ✔ | ✔ | / | | | | | | | |
| eslint-utils | Write file eslint-utils-success | ✔ | ✔ | ✔ | ✔ | ✔ | ✔ | ✔ | ✔ | ✔ | ✔ | ✔ | ✔ | ✔ | ✔ | ✔ | ✔ |
| front-matter | Print 1 | ✘ | ✔ | ✔ | ✔ | ✔ | ✔ | ✔ | ✔ | / | | | | | | | |
| grunt | Returns Date.now | ✔ | ✘ | ✘ | ✔ | ✔ | ✔ | ✔ | ✔ | ✔ | ✔ | ✔ | ✔ | ✔ | ✔ | ✔ | ✔ |
| hot-formula-parser | Write file test | ✔ | ✔ | ✔ | ✔ | ✔ | ✔ | ✔ | ✔ | ✔ | ✔ | ✔ | ✔ | ✔ | ✔ | ✔ | ✔ |
| is-my-json-valid | Execute cat /etc/passwd | ✔ | ✔ | ✔ | ✔ | ✔ | ✔ | ✔ | ✔ | / | | | | | | | |
| jsen | Write file malicious | ✔ | ✘ | ✘ | ✘ | ✘ | ✘ | ✘ | ✘ | / | | | | | | | |
| js-yaml | Returns Date.now | ✔ | ✔ | ✔ | ✔ | ✔ | ✔ | ✔ | ✔ | / | | | | | | | |
| kmc | Write file kmc-success | ✔ | ✔ | ✔ | ✔ | ✔ | ✔ | ✔ | ✔ | ✔ | ✔ | ✔ | ✔ | ✔ | ✔ | ✔ | ✔ |
| marsdb | Write file marsdb-success | ✔ | ✔ | ✔ | ✔ | ✔ | ✔ | ✔ | ✔ | ✔ | ✔ | ✔ | ✔ | ✔ | ✔ | ✔ | ✔ |
| metacalc | Print process | ✘ | ✔ | ✔ | ✔ | ✔ | ✔ | ✔ | ✔ | / | | | | | | | |
| mixin-pro | Print hacked | ✘ | ✔ | ✔ | ✔ | ✔ | ✔ | ✔ | ✔ | ✔ | ✔ | ✔ | ✔ | ✔ | ✔ | ✔ | ✔ |
| m-log | Print injected | ✘ | ✔ | ✔ | ✔ | ✔ | ✔ | ✔ | ✔ | ✔ | ✔ | ✔ | ✔ | ✔ | ✔ | ✔ | ✔ |
| mock2easy | Write mock2easy-success | ✘ | ✔ | ✔ | ✔ | ✔ | ✘ | ✘ | ✘ | ✔ | ✔ | ✔ | ✔ | ✔ | ✔ | ✔ | ✔ |
| modjs | Write modjs-success.txt | ✘ | ✘ | ✘ | ✔ | ✔ | ✘ | ✘ | ✘ | ✔ | ✔ | ✔ | ✔ | ✔ | ✔ | ✔ | ✔ |
| modulify | Print hacked | ✔ | ✔ | ✔ | ✔ | ✔ | ✔ | ✔ | ✔ | ✔ | ✔ | ✔ | ✔ | ✔ | ✔ | ✔ | ✔ |
| mol-proto | Write file mol-proto-success | ✘ | ✔ | ✔ | ✔ | ✔ | ✔ | ✔ | ✔ | ✔ | ✔ | ✔ | ✔ | ✔ | ✔ | ✔ | ✔ |
| mongodb-query-parser | touch test-file | ✔ | ✔ | ✔ | ✔ | ✔ | ✘ | ✘ | ✘ | / | | | | | | | |
| mongo-express | exec calculator | ✔ | ✔ | ✔ | ✔ | ✔ | ✔ | ✔ | ✔ | / | | | | | | | |
| mongoosemask | Print "my evil code was run" | ✘ | ✔ | ✔ | ✔ | ✔ | ✔ | ✔ | ✔ | / | | | | | | | |
| mongo-parse | Write file hacked | ✔ | ✔ | ✔ | ✔ | ✔ | ✔ | ✔ | ✔ | ✔ | ✔ | ✔ | ✔ | ✔ | ✔ | ✔ | ✔ |
| morgan | Write file mongui-success | ✘ | ✔ | ✔ | ✔ | ✔ | ✔ | ✔ | ✔ | ✔ | ✔ | ✔ | ✔ | ✔ | ✔ | ✔ | ✔ |
| morgan-json | Print GLOBAL CTF HIT | ✔ | ✔ | ✔ | ✔ | ✔ | ✔ | ✔ | ✔ | ✔ | ✔ | ✔ | ✔ | ✔ | ✔ | ✔ | ✔ |
| mosc.js | Write file Song | ✔ | ✔ | ✔ | ✔ | ✔ | ✔ | ✔ | ✔ | ✔ | ✔ | ✔ | ✔ | ✔ | ✔ | ✔ | ✔ |
| node-extend | Print 123 | ✘ | ✔ | ✔ | ✔ | ✔ | ✔ | ✔ | ✔ | ✔ | ✔ | ✔ | ✔ | ✔ | ✔ | ✔ | ✔ |
| node-extend | Print 123 | ✘ | ✔ | ✔ | ✔ | ✔ | ✔ | ✔ | ✔ | ✔ | ✔ | ✔ | ✔ | ✔ | ✔ | ✔ | ✔ |
| node-rules.js | Print 123 | ✘ | ✔ | ✔ | ✔ | ✔ | ✔ | ✔ | ✔ | ✔ | ✔ | ✔ | ✔ | ✔ | ✔ | ✔ | ✔ |
| node-serialize | Execute ls | ✘ | ✔ | ✔ | ✔ | ✔ | ✔ | ✔ | ✔ | ✔ | ✔ | ✔ | ✔ | ✔ | ✔ | ✔ | ✔ |
| notevil | Print pwned | ✔ | ✔ | ✔ | ✔ | ✔ | ✔ | ✔ | ✔ | ✔ | ✔ | ✔ | ✔ | ✔ | ✔ | ✔ | ✔ |
| pg | Print process.env | ✘ | ✔ | ✔ | ✔ | ✔ | ✔ | ✔ | ✔ | ✔ | ✔ | ✔ | ✔ | ✔ | ✔ | ✔ | ✔ |
| pixl-class | Print 123 | ✘ | ✔ | ✔ | ✔ | ✔ | ✔ | ✔ | ✔ | ✔ | ✔ | ✔ | ✔ | ✔ | ✔ | ✔ | ✔ |
| post-loader | Print rce | ✔ | ✔ | ✔ | ✔ | ✔ | ✔ | ✔ | ✔ | ✔ | ✔ | ✔ | ✔ | ✔ | ✔ | ✔ | ✔ |
| protojs | Write file protojs-success | ✔ | ✔ | ✔ | ✔ | ✔ | ✔ | ✔ | ✔ | ✔ | ✔ | ✔ | ✔ | ✔ | ✔ | ✔ | ✔ |
| realms-shim | Messed with Object.toString | ✔ | ✔ | ✔ | ✔ | ✔ | ✔ | ✔ | ✔ | ✔ | ✔ | ✔ | ✔ | ✔ | ✔ | ✔ | ✔ |
| reduce-css-calc | Read /etc/passwd | ✔ | ✔ | ✔ | ✔ | ✔ | ✔ | ✔ | ✔ | ✔ | ✔ | ✔ | ✔ | ✔ | ✔ | ✔ | ✔ |
| safe-eval | Return proces | ✔ | ✔ | ✔ | ✔ | ✔ | ✔ | ✔ | ✔ | ✔ | ✔ | ✔ | ✔ | ✔ | ✔ | ✔ | ✔ |
| safer-eval | Print id | ✘ | ✔ | ✔ | ✔ | ✔ | ✔ | ✔ | ✔ | ✔ | ✔ | ✔ | ✔ | ✔ | ✔ | ✔ | ✔ |
| sandbox | Print process.pid | ✘ | ✘ | ✘ | ✔ | ✔ | ✔ | ✔ | ✔ | ✔ | ✘ | ✘ | ✔ | ✔ | ✔ | ✔ | ✔ |
| serialize-javascript | Print 1 | ✘ | ✔ | ✔ | ✔ | ✔ | ✔ | ✔ | ✔ | ✔ | ✔ | ✔ | ✔ | ✔ | ✔ | ✔ | ✔ |
| serialize-to-js | Execute ls | ✘ | ✔ | ✔ | ✔ | ✔ | ✔ | ✔ | ✔ | ✔ | ✔ | ✔ | ✔ | ✔ | ✔ | ✔ | ✔ |
| static-eval | Print hacked | ✔ | ✘ | ✘ | ✔ | ✔ | ✔ | ✔ | ✔ | ✔ | ✔ | ✔ | ✔ | ✔ | ✔ | ✔ | ✔ |
| tar | Overwrite file | ✔ | ✔ | ✔ | ✔ | ✔ | ✔ | ✔ | ✔ | ✔ | ✔ | ✔ | ✔ | ✔ | ✔ | ✔ | ✔ |
| thenify | Write file Song | ✔ | ✔ | ✔ | ✔ | ✔ | ✔ | ✔ | ✔ | / | | | | | | | |
| total.js | Touch HACKED | ✘ | ✔ | ✔ | ✔ | ✔ | ✘ | ✘ | ✘ | / | | | | | | | |
| total4 | Touch HACKED | ✔ | ✘ | ✘ | ✔ | ✔ | ✘ | ✘ | ✘ | / | | | | | | | |
| typed-function | Execute whoami | ✔ | ✔ | ✔ | ✔ | ✔ | ✔ | ✔ | ✔ | / | | | | | | | |
| underscore | touch HELLO | ✘ | ✔ | ✔ | ✔ | ✔ | ✔ | ✔ | ✔ | ✔ | ✔ | ✔ | ✔ | ✔ | ✔ | ✔ | ✔ |
| value-censorship | Access the Function constructor | ✔ | ✔ | ✔ | ✔ | ✔ | ✔ | ✔ | ✔ | ✔ | ✔ | ✔ | ✔ | ✔ | ✔ | ✔ | ✔ |
| vm2 | return process.env | ✘ | ✔ | ✔ | ✔ | ✔ | ✔ | ✔ | ✔ | ✔ | ✔ | ✔ | ✔ | ✔ | ✔ | ✔ | ✔ |
| zhaoyao91-eval-in-vm | return process.env | ✘ | ✔ | ✔ | ✔ | ✔ | ✔ | ✔ | ✔ | ✔ | ✔ | ✔ | ✔ | ✔ | ✔ | ✔ | ✔ |
| mobile-icon-resizer | Print hacked | ✘ | ✘ | ✘ | ✔ | ✔ | ✔ | ✔ | ✔ | ✔ | ✔ | ✔ | ✔ | ✔ | ✔ | ✔ | ✔ |

✘: Exploits are executed; ✔: Exploits are blocked;

## Table 8: Vulnerability payloads.

| Critical Syscall | | JavaScript | C* |
|---|---|---|---|
| Cmd Execution | **exec** | child_process.exec | execve |
| | **fork** | child_process.fork | fork |
| Permission | **setgit** | process.setuid | setuid |
| | **setuid** | process.setgid | setgid |
| Network | **bind** | net.connect | bind |
| | **connect** | dgram.createSocket.bind | connect |
| | **listen** | server.listen | listen |

*: We compile the C code into binary and execute the binary as a shell.

## Table 9: Engine-required system calls.

| Thread Type | Engine-required System Call |
|---|---|
| Main Thread | mprotect; futex; rt_sigaction; munmap; read; fstat; getpid; open; ioctl; rt_sigprocmask; stat; fcntl; writev; epoll_pwait; pread64; dup3; close; write; getcwd; getdents64; rt_sigreturn; brk; shutdown; statx; readlink; madvise; exit_group; epoll_ctl; mmap; |
| Thread Pool | read; futex; openat; socket; statx; open; exit; close; write; rt_sigprocmask; fcntl; getcwd; madvise; munmap; mmap; |

## Table 10: Critical system calls.

| Type | Critical System Call |
|---|---|
| Cmd Execuion | clone; execveat; execve; munmap; fork; |
| Permission | chmod; mprotect; setgid; setreuid; setuid |
| Network | accept4; accept; bind; connect; listen; recvfrom; socket |

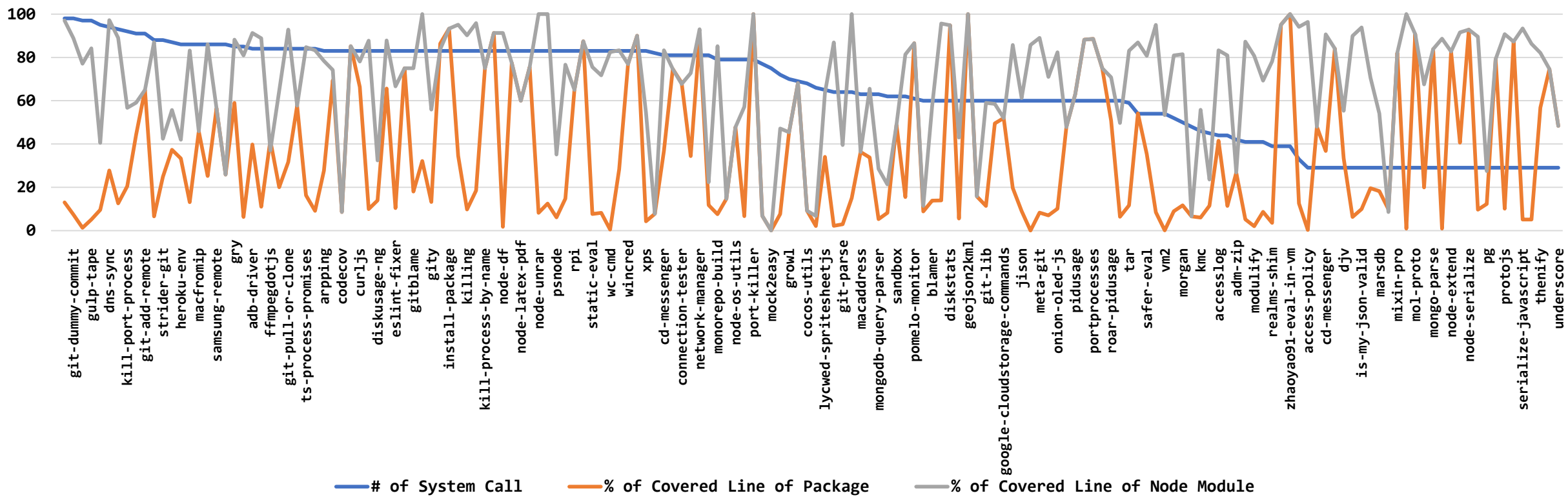

**Figure 9: The relationships between line coverage and syscall number.**